\documentclass[aps,prd,10pt,twocolumn,nofootinbib,superscriptaddress,floatfix]{revtex4-2}

\usepackage{amsmath,amssymb,bm}
\usepackage{graphicx}
\usepackage{booktabs}
\usepackage[hidelinks]{hyperref}
\usepackage{xcolor}
\usepackage{aas_macros}
\hypersetup{colorlinks=true,citecolor=blue,linkcolor=blue,urlcolor=blue}

\newcommand{\Msun}{M_{\odot}}
\newcommand{\Rsun}{R_{\odot}}
\newcommand{\vth}{v_{\rm th}}
\newcommand{\vesc}{v_{\rm esc}}
\newcommand{\mchi}{m_{\chi}}

\begin{document}

\title{Implications of Inelastic Dark Matter for Primordial Dark-Star Evolution:\\ Kinematic Thresholds and Nonthermal Capture}

\author{Qinxun Li}
\affiliation{Department of Physics and Astronomy, University of Utah, Salt Lake City, Utah 84112, USA}
\email{qinxun.li@utah.edu}

\author{Shengyi Liu}
\affiliation{Department of Physics and Jockey Club Institute for Advanced Study,
    The Hong Kong University of Science and Technology, Hong Kong S.A.R., P.R.China}

\date{\today}

\begin{abstract}
    The recent 248-keV nuclear recoil candidate reported by LUX-ZEPLIN has renewed interest in endothermic inelastic dark matter, motivating us to examine its capture in primordial dark stars. Relative to the conventional elastic-capture picture in dark star evolution, endothermic capture adds two qualitative features. First, it opens only after the growing star crosses a compactness threshold, expressed as a kinematic radius $R_{\rm kin}\propto\mu_{\chi}M_\star/\delta$ set by the stellar mass, the dark-matter--nucleus reduced mass and the mass splitting. The accreting growth carries the star through the threshold, and the capture rate turns on quadratically above it. Second, although newly captured particles generically begin on nonthermal bound orbits, endothermic kinematics can keep this normally transient population spatially extended, turning it into a persistent reservoir rather than an intermediate step toward a thermal core. Following complete chains of state-changing collisions, we find that the reservoir compacts sharply and then stalls, because a ground state particle below a compactness-dependent orbital energy has no allowed up-scatter anywhere in the star. The remaining elastic channels are far too slow to bypass this stall, and a percent-level radius contraction reopens it, whether the excited state decays promptly or is long-lived. As a result, inelastic capture does not replenish a thermal annihilation core. The captured population forms an evolving orbital distribution that sets up the co-evolutionary dynamics between the star and the dark matter in the core and the reservoir, which we develop in a companion paper.
\end{abstract}

\maketitle
\raggedbottom

\tableofcontents

\section{Introduction}

Dark stars are hypothetical primordial stars whose luminosity is supplied primarily by dark matter (DM) annihilation rather than nuclear fusion \cite{Spolyar2008,Freese2010SMDS,RindlerDaller2015,FreeseReview2016}. Several high-redshift JWST sources have recently been discussed as possible supermassive dark stars on the basis of their photometry and, in some cases, spectroscopy \cite{Ilie2023JWST,Ilie2025Spectroscopic}. In the conventional scenario, an initially contracted DM distribution supplies the first annihilation heating. Sustained stellar growth then requires continued replenishment of the central DM reservoir, either gravitationally through centrophilic box and chaotic orbits in a nonspherical halo \cite{Freese2010SMDS,2012MNRAS.422.2164I,2010MNRAS.403..525V} or through capture of ambient DM by scattering on stellar nuclei \cite{2008JCAP...11..014F,2008ApJ...677L...1I, 2008MNRAS.390.1655I, 2008ApJ...688L...1Y, 2008PhRvD..78l3510T}. The standard stellar-capture formalism was developed for elastic scattering \cite{PressSpergel1985,1987ApJ...321..560G}. Elastic capture does not place a newly bound particle instantaneously in a thermal core \cite{2009PhRvD..79j3532P}: it first occupies a star-crossing bound orbit and relaxes through subsequent scattering. In conventional capture-supported dark-star evolution this stage is normally treated as a transient preceding a compact captured distribution \cite{2025arXiv251000216T}. However, the instantaneous thermalization approximation can fail when then annihilation heating becomes energetically important before the captured population has thermalized.

Inelastic nuclear scattering provides such a quantitatively different capture channel. A broad class of two-state dark matter models contains a lower state $\chi_1$ and a nearby excited state $\chi_2$, with nuclear scattering dominated by the transition between them \cite{TuckerSmithWeiner2001}. For the endothermic process
\begin{equation}
    \chi_1 + A \rightarrow \chi_2 + A,
    \qquad
    \delta \equiv m_{\chi_2}-m_{\chi_1}>0,
\end{equation}
the collision is possible only above a minimum relative speed set by $\delta$. This makes gravitational capture intrinsically sensitive to the depth of the stellar potential. The same kinematic effect has been studied for the Sun \cite{Nussinov2009,Menon2010,Blennow2016,Blennow2018} and for compact stars \cite{Hooper2010,Bell2018,Alvarez2023,Acevedo2025}. White dwarfs and neutron stars are already sufficiently compact to access large splittings. A primordial dark star is different because its compactness changes substantially as it grows, so the inelastic channel can switch on during the stellar evolution itself.

The recent high-recoil candidate reported by LUX-ZEPLIN (LZ) \cite{LZ2026}---one event near 248 keV, with a global significance of $2.6\sigma$---has renewed interest in endothermic dark matter with splittings of a few hundred keV. Realizations discussed in this context include quasi-Dirac Higgsinos, pseudo-Dirac fermions, inelastic electroweak doublets and multiplets, and inelastic dark-photon dark matter \cite{FreeseTheodosopoulos2026,FanReece2026,DiMauro2026,Visinelli2026,SmirnovGriffithBeacom2026,Yamashita2026}. The quasi-Dirac Higgsino is particularly predictive: for the thermal-relic mass $m_\chi\simeq1.1$ TeV, the off-diagonal $Z$ current fixes the nuclear scattering strength while the neutral-state splitting controls the inelastic kinematics. Solar gravitational acceleration extends endothermic capture beyond terrestrial reach \footnote{Using the IceCube non-observation of high-energy neutrinos from the Sun, Pospelov and Ramani claims a constraint of $\delta>566$ keV, excluding the Higgsino interpretation of the LZ event \cite{PospelovRamani2026}. However, Fan and Reese \cite{FanReece2026} argue that considering the impact of Large Magellen Cloud on the high-speed tail of local dark matter distribution, the required mass splitting can shift to a value that is more consistent with IceCube observations.}. Primordial dark stars access the same kinematics in an evolving gravitational potential: as their central escape speed increases during contraction, an initially forbidden channel can open during stellar evolution, including at splittings above the present solar bound.

\begin{figure*}[!tbp]
    \centering
    \includegraphics[width=\textwidth]{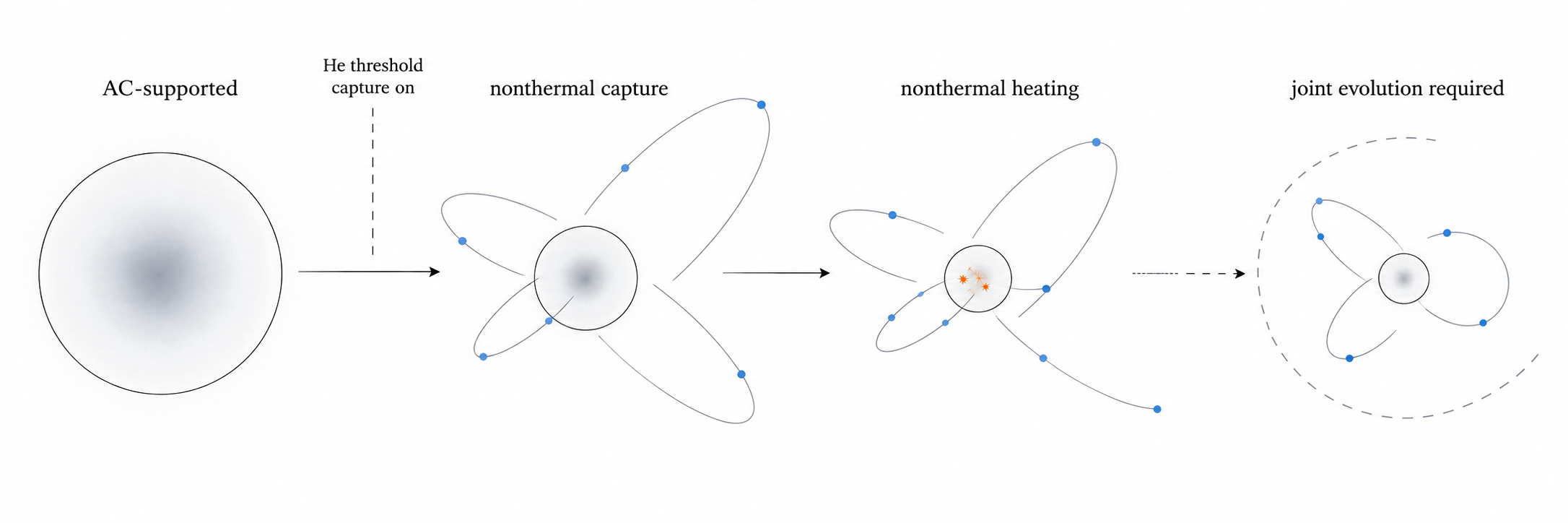}
    \caption{Physical sequence from the kinematic opening of inelastic capture to coupled stellar--dark-matter evolution. An initially AC-supported star crosses the helium capture threshold turns on the inelastic channel. The captured dark matters populate highly eccentric, spatially extended orbits. Their overlap with the pre-existing AC background provides a linear annihilation channel before substantial orbital compaction, while captured--captured annihilation remains inefficient. Later state-changing rescattering can contract part of the reservoir as the potential evolves. A finite stellar response marks the point at which joint stellar dark matter evolution is required. Blue points denote captured DM, gray shading the pre-existing AC background, and orange symbols annihilation heating.}
    \label{fig:illustration}
\end{figure*}

In this work, we study how endothermic capture alters the sequence from first capture to orbital relaxation in a growing dark star. Figure~\ref{fig:illustration} summarizes the physical picture. Once the compactness threshold is crossed, first capture populates spatially extended, highly eccentric orbits that subsequently evolve through state-changing relaxation. Although this geometry suppresses captured--captured annihilation, the same orbits continue to cross the pre-existing adiabatically contracted (AC) dark matter, so the annihilation rate must be computed from the evolving orbital distribution rather than from an assumed thermal core. Further rescattering can later compact part of the stored reservoir as the stellar potential evolves. The central question addressed here is therefore not whether first capture can be nonthermal, but whether inelastic kinematics makes this normally transient stage persistent. The annihilation heating supplied by the resulting reservoir, and its consequences along a supermassive-dark-star growth sequence, are developed in a companion paper.


The paper is organized as follows. Section~\ref{sec:kinematics} derives the endothermic capture kinematics, especially the compactness threshold. Section~\ref{sec:captureonset} computes the near-threshold capture rate and locates the threshold crossing along supermassive-dark-star growth tracks. Section~\ref{sec:postcapture} evolves the phase space populated by first capture and the follow-up state-changing relaxation. Section~\ref{sec:stall} simulates the resulting collision chains to their kinematic endpoint. The physical interpretation and scope of the result are discussed in Section~\ref{sec:discussion}.

\section{Endothermic capture and the compactness threshold}
\label{sec:kinematics}

We first derive the conditions under which an incoming particle can scatter and become bound. The interaction fixes the scattering normalization, while the $n=3$ stellar potential turns the inelastic velocity threshold into a compactness condition.

\subsection{Interaction normalization and mass splitting}

We take the interaction to be an off diagonal neutral current between two nearby Majorana states. A nearly pure Higgsino provides a concrete electroweak realization: electroweak symmetry breaking splits the neutral Dirac Higgsino into two nearby mass eigenstates, $\chi_1$ and $\chi_2$, and the $Z$ current is off diagonal in this basis \cite{2015PhRvD..91e5035N, FreeseTheodosopoulos2026,GrahamRamaniWong2025}. Schematically, up to a Majorana phase convention,
\begin{equation}
    {\cal L}_{\rm NC}\supset
    i\frac{g}{2c_W}Z_\mu\,
    \overline{\chi_1}\gamma^\mu\chi_2,
    \label{eq:higgsinoZ}
\end{equation}
so tree level $Z$ exchange produces $\chi_1 A\leftrightarrow\chi_2 A$ scattering. The corresponding zero momentum nucleon cross section is fixed by the weak interaction,
\begin{equation}
    \sigma^0_{\widetilde H N}
    =
    \frac{G_F^2\mu_N^2}{2\pi}
    \simeq7.43\times10^{-39}\ {\rm cm^2},
    \label{eq:higgsinoN}
\end{equation}
where we use the full vector-like off-diagonal normalization of Pospelov and Ramani \cite{PospelovRamani2026}. For the H/He momentum transfers relevant here, $q^2\ll m_Z^2$, so $Z$ exchange is accurately described by the contact interaction used below.

The Higgsino interpretation also fixes the thermal mass scale. Standard thermal freeze-out gives $m_\chi\simeq1.1$ TeV \cite{HiggsinoReview2018}. Ground-state annihilation proceeds efficiently through electroweak channels including $W^+W^-$ and $ZZ$, while the excited state decays rapidly on orbital timescales. We evaluate the orbital dynamics at
\begin{equation}
    m_\chi=1.1~{\rm TeV},
    \qquad
    \delta_{\rm ref}=350~{\rm keV},
    \label{eq:higgsinoBenchmark}
\end{equation}
using $\delta_{\rm ref}=350$ keV as a reference for the compactness-dependent orbital calculations. Results are also quoted at $\delta=566$ keV, corresponding to the current solar Higgsino bound. The weak interaction scattering normalization is independent of $\delta$, so the threshold location can be varied separately from the scattering strength; its dependence across the quasi-Dirac regime is given in Sec.~\ref{sec:tracks}.

\subsection{Endothermic capture kinematics}

Let $\rho_{\chi,\infty}$ denote the unbound far-field density that supplies capture; $\rho_{\rm AC}(r)$ denotes the bound background used for annihilation overlap. Consider an unbound DM particle with speed $u$ far from the star. At radius $r$ its speed before scattering is
\begin{equation}
    w^2(r)=u^2+\vesc^2(r).
    \label{eq:w}
\end{equation}
For a target nucleus of mass $m_A$, define
\begin{equation}
    \mu_{\chi A}=\frac{\mchi m_A}{\mchi+m_A}.
\end{equation}
The endothermic transition is kinematically allowed only if
\begin{equation}
    \frac{1}{2}\mu_{\chi A}w^2(r)>\delta .
    \label{eq:inelastic_threshold}
\end{equation}
Equivalently, the minimum relative speed is
\begin{equation}
    \vth,A=\left(\frac{2\delta}{\mu_{\chi A}}\right)^{1/2}.
    \label{eq:vth}
\end{equation}

A particle is gravitationally captured if its energy after the collision is negative. If $E_R$ is the nuclear recoil energy, the DM loses an energy $E_R+\delta$, and capture requires
\begin{equation}
    E_R+\delta>\frac12\mchi u^2.
    \label{eq:binding}
\end{equation}
A useful limit follows immediately. Define
\begin{equation}
    u_{\rm auto}\equiv\left(\frac{2\delta}{\mchi}\right)^{1/2}.
    \label{eq:uauto}
\end{equation}
For $u<u_{\rm auto}$, every kinematically allowed endothermic scatter satisfies Eq.~(\ref{eq:binding}) because $E_R\ge0$. For $\mchi=1.1$ TeV and $\delta=350$ keV,
\begin{equation}
    u_{\rm auto}\simeq239~{\rm km\,s^{-1}},
\end{equation}
which is much larger than the velocity dispersion expected in primordial minihalos. Consequently, the leading suppression in this regime arises from Eq.~(\ref{eq:inelastic_threshold}), rather than from an insufficient energy loss after a collision.

The kinematically allowed nuclear-recoil interval is \cite{Blennow2016}
\begin{align}
    E_{\pm}(w)
     & =
    \frac{\mu_{\chi A}^{2}}{m_A}w^{2}
    \left(
    1\pm\sqrt{1-\frac{2\delta}{\mu_{\chi A}w^{2}}}
    \right)
    -\frac{\mu_{\chi A}}{m_A}\delta .
    \label{eq:recoil_bounds}
\end{align}
Capture requires
\begin{equation}
    E_R>E_{\rm capt}(u)
    \equiv \frac12m_\chi u^2-\delta .
    \label{eq:Ecapture}
\end{equation}
For a contact interaction normalized by the zero-momentum nuclear cross section $\sigma_A^0$, the differential cross section within Eq.~(\ref{eq:recoil_bounds}) is
\begin{equation}
    \frac{d\sigma_A}{dE_R}
    =
    \frac{m_A\sigma_A^0}
    {2\mu_{\chi A}^{2}w^{2}}
    F_A^2(E_R).
    \label{eq:dsigdER}
\end{equation}
With the differential normalization used here, the endothermic phase-space suppression is carried by the shrinking allowed recoil interval rather than by an additional multiplicative factor in $d\sigma_A/dE_R$ \cite{Blennow2016}. The cross section for a first scatter that produces a bound orbit is therefore
\begin{equation}
    \begin{aligned}
        \sigma_{{\rm cap},A}(w,u)
                    & =
        \int_{E_{\rm low}}^{E_+(w)}
        dE_R\,\frac{d\sigma_A}{dE_R},
        \\
        E_{\rm low} & =\max[E_-(w),E_{\rm capt}(u),0],
    \end{aligned}
    \label{eq:sigmacap}
\end{equation}
with the integral set to zero for $E_{\rm low}\ge E_+$. For a normalized stellar-frame speed distribution $f(u)$, the exact optically thin single-scatter capture rate takes the standard stellar-capture form \cite{PressSpergel1985,Gould1987},
\begin{equation}
    \frac{dC_A}{dV}
    =
    \frac{\rho_{\chi,\infty}}{m_\chi}n_A(r)
    \int_0^\infty du\,\frac{f(u)}{u}
    w^2\sigma_{{\rm cap},A}(w,u).
    \label{eq:exactcapture}
\end{equation}

\subsection{Compactness threshold in an \texorpdfstring{$n=3$}{n=3} dark star}
\label{sec:polytrope}

The supermassive dark-star models of Rindler-Daller et al. are well described by $n=3$ polytropes \cite{RindlerDaller2015}, appropriate to their radiation-pressure-dominated structure. Let $\theta(\xi)$ satisfy the Lane--Emden equation,
\begin{equation}
    \frac{1}{\xi^2}\frac{d}{d\xi}
    \left(\xi^2\frac{d\theta}{d\xi}\right)
    =-\theta^3,
\end{equation}
with surface $\theta(\xi_1)=0$. Numerically,
\begin{align}
    \xi_1           & =6.8968486,
                    & -\xi_1^2\theta'_1 & =2.0182360, \\
    -\xi_1\theta'_1 & =0.2926316.
\end{align}
The escape speed can be written as
\begin{equation}
    \vesc^2(\xi)
    =
    \frac{2GM_\star}{R_\star}
    \left[
        1+\frac{\theta(\xi)}{-\xi_1\theta'_1}
        \right].
    \label{eq:vesc_profile}
\end{equation}
At the center,
\begin{equation}
    \vesc^2(0)=2\eta_3\frac{GM_\star}{R_\star},
    \qquad
    \eta_3
    =1+\frac{1}{-\xi_1\theta'_1}
    =4.4172658 .
    \label{eq:eta3}
\end{equation}

The existence of any kinematically allowed region inside the star requires
\begin{equation}
    \delta<
    \eta_3\mu_{\chi A}\frac{GM_\star}{R_\star}.
\end{equation}
We therefore define the species-dependent kinematic radius
\begin{equation}
    R_{\rm kin,A}
    =
    \eta_3\frac{\mu_{\chi A}c^2}{\delta}
    \frac{GM_\star}{c^2}.
    \label{eq:Rkin}
\end{equation}
For stationary targets and negligible halo speed, $R_\star>R_{\rm kin,A}$ forbids inelastic scattering on species $A$ throughout the star. Finite target velocities round this stationary-target boundary.

For $\mchi=1.1$ TeV and $\delta=350$ keV,
\begin{align}
    R_{\rm kin,H}
     & \simeq
    0.117~{\rm AU}
    \left(\frac{M_\star}{10^3\Msun}\right), \\
    R_{\rm kin,He}
     & \simeq
    0.463~{\rm AU}
    \left(\frac{M_\star}{10^3\Msun}\right).
    \label{eq:Rkin_benchmark}
\end{align}
The weak dependence on $\mchi$ in the heavy-DM regime is contained in the reduced mass.

Endothermic scattering therefore introduces a species-dependent compactness threshold. For $\delta=350$ keV, the helium channel opens at a radius about four times larger than the hydrogen channel. We next determine how rapidly the capture rate grows once this threshold is crossed.

\section{Capture onset in primordial dark stars}
\label{sec:captureonset}

Near threshold, halo and target velocities broaden the opening, while composition and finite nuclear-size effects set its normalization. These effects determine how the helium channel turns on and where that opening occurs along the small-minihalo (SMH) and large-minihalo (LMH) growth tracks.

\subsection{Near-threshold capture rate}
\label{sec:rate}

We next consider the low-$u$, optically thin limit, in which every allowed scatter captures the incoming particle. For a contact interaction normalized by an elastic-limit cross section $\sigma_A$, the inelastic phase-space factor contributes
\begin{equation}
    \beta_A(w)
    =
    \left(1-\frac{2\delta}{\mu_{\chi A}w^2}\right)^{1/2}.
\end{equation}
Neglecting terms of order $u^2/\vesc^2$, the capture rate can be organized as
\begin{equation}
    C_A\simeq
    \frac{\rho_{\chi,\infty}}{\mchi}
    \left\langle\frac1u\right\rangle
    \sigma_A N_A
    \vesc^2(0)
    J_3(\lambda_A),
    \label{eq:CA}
\end{equation}
where
\begin{equation}
    \lambda_A
    \equiv
    \frac{2\delta}{\mu_{\chi A}\vesc^2(0)}
    =
    \frac{R_\star}{R_{\rm kin,A}}.
\end{equation}
With
\begin{equation}
    g(\xi)\equiv
    \frac{\vesc^2(\xi)}{\vesc^2(0)}
    =
    \frac{1+\alpha_3\theta(\xi)}{\eta_3},
    \qquad
    \alpha_3\equiv\frac{1}{-\xi_1\theta'_1},
\end{equation}
the point-nucleus structure integral is
\begin{equation}
    J_3(\lambda)=
    \frac{
        \int d\xi\,\xi^2\theta^3
        g(\xi)
        \sqrt{1-\lambda/g(\xi)}
        \Theta[g(\xi)-\lambda]
    }{
        \int d\xi\,\xi^2\theta^3
    }.
    \label{eq:J3}
\end{equation}
Numerical integration gives
\begin{equation}
    J_3(0)=0.679153.
\end{equation}
Near the kinematic boundary,
\begin{equation}
    J_3(\lambda)
    =
    2.101(1-\lambda)^2
    +{\cal O}\!\left[(1-\lambda)^3\right].
    \label{eq:J3asymptotic}
\end{equation}
Thus, the inelastic capture rate exhibits a steep but continuous turn-on,
\begin{equation}
    C_A\propto
    \left(1-\frac{R_\star}{R_{\rm kin,A}}\right)^2,
    \qquad
    R_\star\rightarrow R_{\rm kin,A}^{-}.
    \label{eq:turnon}
\end{equation}

\begin{figure*}[!tbp]
    \centering
    \includegraphics[width=\textwidth]{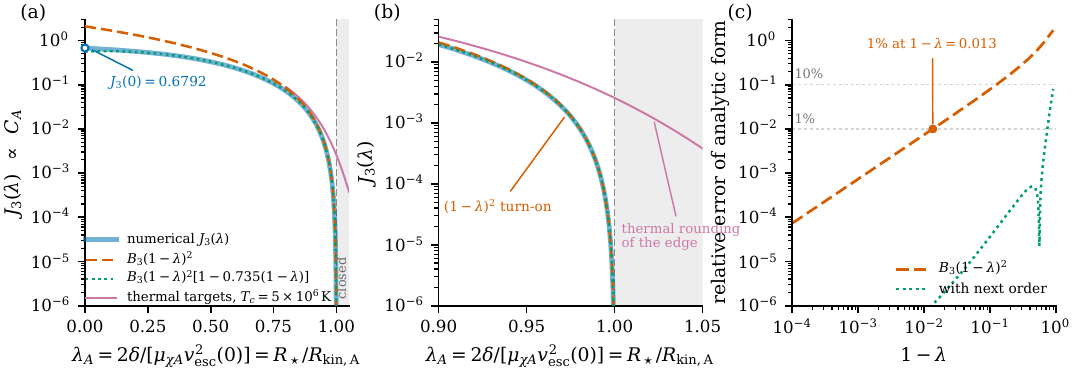}
    \caption{Dimensionless capture structure integral $J_3(\lambda)$,
        Eq.~(\ref{eq:J3}), compared with its near-threshold analytic limit. Because
        $C_A\propto J_3(\lambda_A)$ at fixed $\lambda$-independent prefactor in
        Eq.~(\ref{eq:CA}), the ordinate is the normalized capture rate.
        (a) Full range: numerical quadrature over the $n=3$ Lane--Emden profile
        (thick line) against $B_3(1-\lambda)^2$ (Eq.~\ref{eq:J3asymptotic}, dashed)
        and the same form with the next-order term (Eq.~\ref{eq:J3nextorder},
        dotted); the open circle marks $J_3(0)=0.679153$. (b) Neighborhood of the
        kinematic boundary. The point-nucleus thermal target result of
        Eq.~(\ref{eq:thermal_closed}) at $T_c=5\times10^6$ K rounds the
        stationary-target edge and leaks past $\lambda=1$; the halo-velocity
        rounding of Eq.~(\ref{eq:finiteu_boundary}), with
        $\epsilon_{\rm He}\lesssim4\times10^{-5}$, acts on a scale three orders of
        magnitude narrower and is not resolved here. (c) Relative error of the two
        analytic forms as a function of $1-\lambda$.}
    \label{fig:J3threshold}
\end{figure*}

As shown in the left and middle panels of Figure~\ref{fig:J3threshold},the quadratic approximation in Eq.~\ref{eq:J3asymptotic} works quite good near the threshold, reproducing $J_3$ to better than $1\%$ for
$1-\lambda<0.014$ and to $10\%$ for $1-\lambda<0.124$. Retaining the next
order,
\begin{equation}
    J_3(\lambda)=B_3(1-\lambda)^2
    \left[1-0.735(1-\lambda)+{\cal O}\!\left((1-\lambda)^2\right)\right],
    \label{eq:J3nextorder}
\end{equation}
where
\begin{equation}
    B_3=
    \frac{\pi}
    {16A_3^{3/2}(-\xi_1^2\theta'_1)}
    =2.101
\end{equation}
extends the $1\%$ range to $\lambda\gtrsim0.27$, as shown in the right panel of Figure~\ref{fig:J3threshold}.

Finite halo velocities smooth the nominal boundary at $\lambda_A=1$. This effect admits a simple analytic form in the same near-threshold expansion. Near the center of an $n=3$ polytrope,
\begin{equation}
    g(\xi)=1-A_3\xi^2+{\cal O}(\xi^4),
    \qquad
    A_3=\frac{\alpha_3}{6\eta_3}=0.128936.
\end{equation}
Define
\begin{equation}
    \Delta\equiv1-\lambda_A,
    \qquad
    \epsilon_A\equiv
    \left(\frac{\sigma_u}{v_{{\rm th},A}}\right)^2,
\end{equation}
where $\sigma_u$ is the one-dimensional dispersion of a Maxwellian halo. In the automatic-binding regime, velocity averaging gives
\begin{equation}
    J_3(\Delta,\epsilon_A)
    \simeq
    B_3
    \begin{cases}
        \Delta^2+4\Delta\epsilon_A+8\epsilon_A^2,
         & \Delta\ge0, \\[2pt]
        8\epsilon_A^2
        \exp[\Delta/(2\epsilon_A)],
         & \Delta<0,
    \end{cases}
    \label{eq:finiteu_boundary}
\end{equation}
The coefficients 4 and 8 arise from the velocity moments under the $f(u)/u$ capture weighting: after normalization, $\langle u^2\rangle=2\sigma_u^2$ and $\langle u^4\rangle=8\sigma_u^4$. The zero-velocity result in Eq.~(\ref{eq:J3asymptotic}) is recovered as $\epsilon_A\rightarrow0$.

For $m_\chi=1.1$ TeV and $\delta=350$ keV, $v_{{\rm th,He}}\simeq4115~{\rm km\,s^{-1}}$. Thus $\epsilon_{\rm He}=1.5\times10^{-6}$ and $3.7\times10^{-5}$ for $\sigma_u=5$ and $25~{\rm km\,s^{-1}}$, respectively. Direct evaluation of Eq.~(\ref{eq:exactcapture}) shows that the zero-velocity approximation is accurate to better than $0.8\%$ at $\lambda_{\rm He}\le0.98$ and to $1.5\%$ at $\lambda_{\rm He}=0.99$ for $\sigma_u=25~{\rm km\,s^{-1}}$, and is substantially more accurate for smaller velocity dispersions. Leakage above the nominal threshold is exponentially confined to a fractional compactness interval of order $\epsilon_{\rm He}$. The halo-velocity rounding is therefore negligible and not shown in Figure~\ref{fig:J3threshold}.

\subsection{Thermal motion of stellar targets}
\label{sec:thermal_targets}

The thermal motion of stellar nuclei provides a parametrically larger rounding of the threshold than the primordial-halo velocity dispersion. Let $\bm v_A$ denote the target velocity and define
\begin{equation}
    \bm s=\bm w-\bm v_A,\qquad
    \beta(s)=
    \sqrt{1-\frac{\vth,A^2}{s^2}}\,
    \Theta(s-\vth,A).
    \label{eq:thermalbeta}
\end{equation}
Neglecting corrections of relative order $\delta/\mchi$ in the inertial masses, the center-of-mass velocity and nuclear mass fraction are
\begin{equation}
    \bm V=
    \frac{\mchi\bm w+m_A\bm v_A}{\mchi+m_A},
    \qquad
    \kappa_A=\frac{m_A}{\mchi+m_A}.
\end{equation}
The outgoing DM velocity can be parameterized as
\begin{equation}
    \bm w'=\bm V+\kappa_A s\beta\,\hat{\bm n}'.
\end{equation}
The condition $w'<\vesc$ may therefore be integrated over the isotropic outgoing direction. Defining
\begin{equation}
    x_c=
    \frac{\vesc^2-V^2-\kappa_A^2s^2\beta^2}
    {2\kappa_AVs\beta},
\end{equation}
the point-nucleus capture probability is
\begin{equation}
    P_{\rm cap}=
    \begin{cases}
        0,         & x_c\le-1, \\
        (1+x_c)/2, & -1<x_c<1, \\
        1,         & x_c\ge1.
    \end{cases}
    \label{eq:PcapT}
\end{equation}
For Maxwellian targets with one-dimensional thermal dispersion
\begin{equation}
    a_A=\sqrt{\frac{k_BT}{m_A}},
\end{equation}
the point-nucleus local kernel is
\begin{equation}
    K_A(w,T)=
    \int d^3v_A\,f_A(\bm v_A;T)\,
    s\,\beta(s)P_{\rm cap}.
    \label{eq:Kthermal}
\end{equation}
Finite nuclear size is included by replacing
$P_{\rm cap}$ in Eq.~(\ref{eq:Kthermal}) with
\begin{equation}
    \int\frac{d\Omega'}{4\pi}\,
    F_W^2\!\left(\mu_{\chi A}|\bm s-s\beta\hat{\bm n}'|\right)
    \Theta(\vesc^2-|\bm w'|^2).
    \label{eq:thermalWeak}
\end{equation}
Thus the momentum transfer is evaluated in the relative motion while
binding is tested in the stellar frame. The stationary-target recoil
integral is recovered as $T\rightarrow0$.

To quantify the broadening we adopt the $n=3$ polytropic temperature profile $T(\xi)=T_c\theta(\xi)$ and define
\begin{equation}
    J_{3,T}=
    \frac{\int d\xi\,\xi^2\theta^3
    w(\xi)K_A[w(\xi),T(\xi)]/\vesc^2(0)}
    {\int d\xi\,\xi^2\theta^3}.
    \label{eq:J3T}
\end{equation}
For $\mchi=1.1$ TeV and $\delta=350$ keV, including the helium weak form factor
specified in Sec.~\ref{sec:heliumresponse}, the contribution at the nominal stationary-target boundary is
\begin{align}
    J_{3,T}(1,2\times10^6{\rm K}) & =9.0\times10^{-4},  \\
    J_{3,T}(1,5\times10^6{\rm K}) & =2.16\times10^{-3}, \\
    J_{3,T}(1,7\times10^6{\rm K}) & =2.96\times10^{-3}.
    \label{eq:J3Tnumbers}
\end{align}
At $\lambda=0.9$, the corresponding enhancement relative to stationary targets is approximately $6\%$, $15\%$, and $21\%$. Once the channel is well open, the thermal correction rapidly becomes negligible, as shown in Figure~\ref{fig:J3threshold}.

Near the boundary, let $t_A=a_{A,c}/\vth,A$ and $\epsilon=1-\lambda$. A central expansion of the point-nucleus kernel gives
\begin{equation}
    J_{3,T}\simeq
    2.10\left\langle(\epsilon+Z)_+^2\right\rangle,
    \qquad
    Z\sim{\cal N}(0,4t_A^2),
    \label{eq:thermal_asym}
\end{equation}
or
\begin{multline}
    J_{3,T}\simeq2.10\Bigg[
        (\epsilon^2+4t_A^2)
        \Phi\!\left(\frac{\epsilon}{2t_A}\right)\\
        +2\epsilon t_A
        \phi\!\left(\frac{\epsilon}{2t_A}\right)
        \Bigg].
    \label{eq:thermal_closed}
\end{multline}
In particular, $J_{3,T}(1)\rightarrow4.20t_A^2$ in the asymptotic central limit. The leading weak-form-factor correction multiplies these central expressions by $F_W^2(q_{\rm th})$.

The central temperatures of the SMH and LMH sequences at the stationary helium crossings is $T_c\simeq 10.8\times10^6$ K and $8.6\times10^6$ K \footnote{The $m_\chi=1000$ GeV sequences used here do not tabulate $T_c$, so we reconstruct the central temperature from their $M_\star$--$R_\star$ tracks assuming the $n=3$. For an $n=3$ polytrope, $\rho_c/\bar\rho=54.18$ and $P_c=\pi G(R_\star/\xi_1)^2\rho_c^2$; solving $P_c=\rho_c k_B T_c/(\mu m_p)+aT_c^4/3$ for a fully ionized primordial mixture ($\mu\simeq0.588$). As a cross-check, the same reconstruction reproduces the central temperatures tabulated for the $100$ GeV sequences of Ref.~\cite{RindlerDaller2015} to within about $7\%$ over the relevant mass range.}, respectively.  At the $566$ keV crossings the corresponding values are $1.28\times10^7$ K and $1.06\times10^7$ K. The thermal broadening is equivalent to a zero-temperature penetration $1-\lambda\simeq0.04$--$0.05$ at the stationary boundary. The stationary boundary therefore remains a convenient measure of where the thermal turn-on occurs. Numerical details are given in Appendix~\ref{app:kernels}. Over the relevant thermal phase space, the angular integration gives $P_{\rm cap}=1$ to numerical precision, so target motion broadens the scattering threshold without changing the result that every allowed first scatter is gravitationally binding.

\subsection{Target hierarchy and helium weak form factor}
\label{sec:composition}
\label{sec:heliumresponse}

Primordial dark stars are composed predominantly of hydrogen and helium. For reference, an isoscalar spin-independent contact interaction obeys
\begin{equation}
    \sigma_A
    =
    \sigma_p A^2
    \left(\frac{\mu_{\chi A}}{\mu_{\chi p}}\right)^2 ,
    \label{eq:SIscale}
\end{equation}
which gives $\sigma_{\rm He}/\sigma_p\simeq251$ for TeV-scale dark matter. With primordial mass fractions $X=0.76$ and $Y=0.24$,
\begin{equation}
    \frac{N_{\rm He}}{N_H}
    =
    \frac{Y/4}{X}
    \simeq0.079,
\end{equation}
so helium already exceeds hydrogen by a factor $\simeq20$ in the fully open isoscalar rate.

The Higgsino hierarchy is stronger because the vector neutral current is predominantly neutron coupled. For a nucleus with mass number $A$ and charge $Z$,
\begin{equation}
    \sigma^0_{\widetilde H A}
    =
    \frac{G_F^2\mu_{\chi A}^2}{2\pi}
    Q_A^2,
    \qquad
    Q_A=(A-Z)-(1-4\sin^2\theta_W)Z .
    \label{eq:higgsinoNucleus}
\end{equation}
For ${}^4$He, $Q_{\rm He}=1.85$ and
\begin{equation}
    \sigma_{\rm He}^0\simeq3.98\times10^{-37}\ {\rm cm^2},
\end{equation}
whereas the proton weak charge suppresses hydrogen to $\sigma_H^0\simeq4.0\times10^{-41}\ {\rm cm^2}$. Including primordial abundances,
\begin{equation}
    \frac{N_{\rm He}\sigma_{\rm He}}
    {N_H\sigma_H}
    \simeq7.5\times10^2 .
    \label{eq:HiggsinoHeDominance}
\end{equation}
The kinematic threshold reinforces this hierarchy. For $m_\chi=1.1$ TeV and $\delta=350$ keV,
\begin{equation}
    v_{\rm th,H}\simeq8.2\times10^3~{\rm km\,s^{-1}},
    \qquad
    v_{\rm th,He}\simeq4.1\times10^3~{\rm km\,s^{-1}}.
\end{equation}
Helium therefore dominates the capture rate over the compactness range studied here.

The $J^\pi=0^+$ helium ground state selects the coherent vector form factor. Writing $c_p=1-4\sin^2\theta_W$, we normalize it as
\begin{equation}
    F_W(q)=\frac{NF_n(q)-c_pZF_p(q)}{N-c_pZ},\qquad F_W(0)=1.
\end{equation}
In the charge-symmetric, zero-strangeness one-body limit, the normalized weak and electromagnetic shapes of ${}^4$He coincide \cite{Aniol2006}; we therefore use the empirical charge distribution of de Vries \textit{et al.}\ \cite{DeVries1987}. Its sum-of-Gaussians (SOG) implementation and numerical checks are given in Appendix~\ref{app:kernels}.

Finite nuclear size changes the capture normalization but not the stationary threshold. At opening, $q_{\rm th}=\sqrt{2\mu_{\chi\mathrm{He}}\delta}$ and the near-threshold rate becomes
\begin{equation}
    J_{3,W}(\lambda)=2.10134 F_W^2(q_{\rm th})(1-\lambda)^2
    +\mathcal O[(1-\lambda)^3].
\end{equation}
For $\delta=350$ and $566$ keV, $F_W^2(q_{\rm th})=0.939$ and $0.904$, respectively. Thus the helium weak form factor preserves both the compactness threshold and the quadratic opening law while modestly suppressing the rate. The same finite-$q$ form-factor weighting is used for first-capture sampling and subsequent scattering; Appendix~\ref{app:kernels} gives the full splitting dependence and its effect on the post-capture orbital calculations.

Having fixed the dominant target species and its weak form factor,
we can now ask where the compactness threshold is encountered during stellar
growth.

\subsection{Threshold crossing during stellar growth}
\label{sec:tracks}

To estimate where the threshold is crossed during dark-star growth, we compare Eq.~(\ref{eq:Rkin}) with the SMH and LMH MESA~\cite{Paxton2011,Paxton2013} sequences of Rindler-Daller et al.~\cite{RindlerDaller2015}. Their accretion rates are $\dot M=10^{-3}\Msun\,{\rm yr^{-1}}$ for SMH and $10^{-1}\Msun\,{\rm yr^{-1}}$ for LMH. Both sequences include extended adiabatic contraction but omit scattering capture, so their radius--mass relations locate the first kinematic opening before capture occurs.

\begin{figure}[!tbp]
    \centering
    \includegraphics[width=\columnwidth]{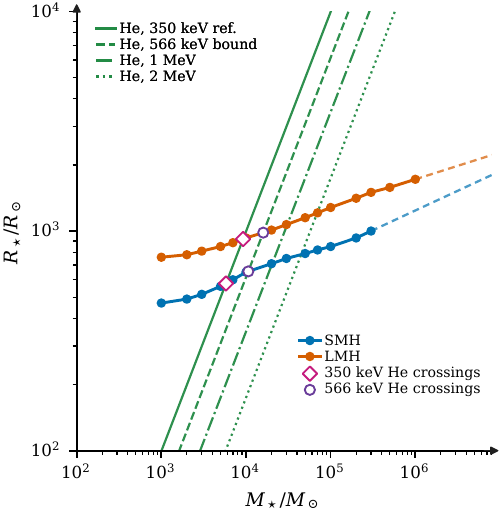}
    \caption{SMH and LMH stellar radius tracks for $m_\chi=1000$ GeV from Figure~9 of Ref.~\cite{RindlerDaller2015}, compared with stationary helium capture thresholds for a $1.1$-TeV Higgsino and $\delta=350$ keV, $566$ keV (solar bound), $1$ MeV, and $2$ MeV. Points are digitized MESA models and solid segments connect adjacent points; dashed extensions continue the local high-mass trend beyond the digitized sequence. Below a threshold line the helium channel is open at the stellar center. Crossings with the 350-keV and 566-keV threshold lines are marked.}
    \label{fig:stellarThresholdTracks}
\end{figure}

For consistency with the $m_\chi=1.1$ TeV Higgsino used in the capture calculation, we use the $m_\chi=1000$ GeV stellar sequences shown in Figure~9 of Ref.~\cite{RindlerDaller2015}. These tracks are substantially more compact than the 10 GeV and 100 GeV models, giving
\begin{equation}
    \begin{split}
        R_\star^{\rm SMH}(10^4\Msun) & \simeq 6.5\times10^2\Rsun, \\
        R_\star^{\rm SMH}(10^5\Msun) & \simeq 8.5\times10^2\Rsun, \\
        R_\star^{\rm LMH}(10^4\Msun) & \simeq 9.3\times10^2\Rsun, \\
        R_\star^{\rm LMH}(10^5\Msun) & \simeq 1.3\times10^3\Rsun.
    \end{split}
\end{equation}
As shown in Figure~\ref{fig:stellarThresholdTracks}, the helium threshold for $m_\chi=1.1$ TeV and $\delta=350$ keV then intersects the tracks at
\begin{equation}
    M_{\rm kin}^{\rm SMH}\simeq5.8\times10^3\Msun,
    \qquad
    M_{\rm kin}^{\rm LMH}\simeq9.2\times10^3\Msun.
    \label{eq:Mkin}
\end{equation}
At the current solar Higgsino bound, $\delta=566$ keV, the corresponding crossings move to $1.1\times10^4\Msun$ and $1.6\times10^4\Msun$, respectively. Even MeV-scale splittings remain kinematically accessible well within the supermassive-dark-star mass range.

\begin{table}[!tbp]
    \caption{Helium threshold masses obtained by logarithmic interpolation of the digitized $m_\chi=1000$ GeV SMH and LMH radius tracks shown in Figure~9 of Ref.~\cite{RindlerDaller2015}.  The 5-MeV row approaches the regime where general-relativistic instability may become relevant and is included to show the kinematic reach \cite{2024A&A...689A.202H, 2025arXiv251108578F}.}
    \label{tab:kinreach}
    \begin{ruledtabular}
        \begin{tabular}{ccc}
            $\delta$ & $M_{\rm kin}^{\rm SMH}/\Msun$ & $M_{\rm kin}^{\rm LMH}/\Msun$ \\
            \hline
            350 keV  & $5.8\times10^3$               & $9.2\times10^3$               \\
            566 keV  & $1.1\times10^4$               & $1.6\times10^4$               \\
            1 MeV    & $2.0\times10^4$               & $3.1\times10^4$               \\
            2 MeV    & $4.5\times10^4$               & $6.9\times10^4$               \\
            5 MeV    & $1.3\times10^5$               & $2.0\times10^5$               \\
        \end{tabular}
    \end{ruledtabular}
\end{table}

Fitting the evolution tracks with power law mass-radius relation
\begin{equation}
    R_{\star}\propto M_{\star}^\alpha,
    \label{eq:MkinScaling}
\end{equation}
we get approximately $\lambda=R_\star/R_{\rm kin}\propto M_\star^{-0.86}$ for SMH and $M_\star^{-0.87}$ for LMH. The negative exponent implies that by accreting mass, a star can naturally approach the compactness threshold and turn on the inelastic capture channel. At the 350-keV crossings of Eq.~(\ref{eq:Mkin}), the corresponding accretion times are about $5.8\times10^6$ yr (SMH) and $9.2\times10^4$ yr (LMH). The much longer SMH clock makes accumulation of a nonthermal reservoir particularly relevant once the channel opens.

\begin{figure}[t]
    \centering
    \includegraphics[width=\columnwidth]
    {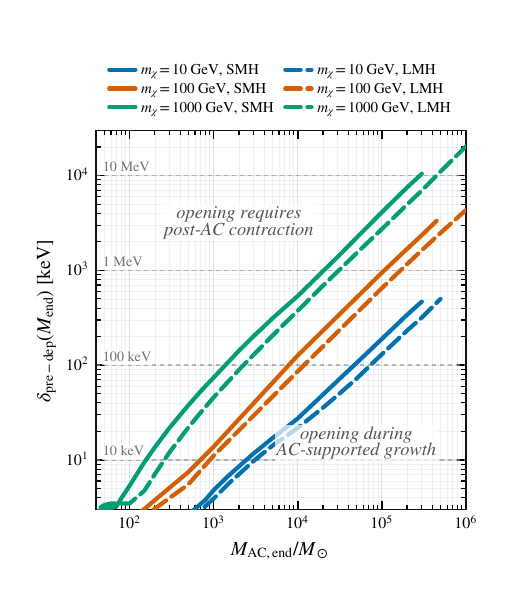}
    \caption{
    Maximum helium endothermic splitting that becomes kinematically
    accessible before termination of the AC-supported growth phase.
    The curves are evaluated along the SMH and LMH sequences of
    Rindler-Daller et al.~\cite{RindlerDaller2015} for
    $m_\chi=10$, $100$, and $1000~{\rm GeV}$, with the capture-particle
    mass taken to be the same as that defining each stellar sequence.
    For a putative AC-support termination mass $M_{\rm AC,end}$,
    Eq.~(\ref{eq:delta_predep}) gives the cumulative maximum of the
    stationary-target helium threshold attained earlier along the
    sequence.  For each curve, splittings below the boundary open
    during AC-supported growth; splittings above it require further
    post-AC contraction.  The short low-mass plateaus in the
    $1~{\rm TeV}$ curves arise from the nonmonotonic radius feature in
    the corresponding stellar models.
    }
    \label{fig:ac_accessibility}
\end{figure}

However, the mass of a dark star can not be increased indefinitely along the $M_\star$--$R_\star$ relation of the AC-supported sequence.
If annihilation depletes the contracted dark-matter
reservoir before the formal crossing mass is reached, the star leaves
the extended-AC branch and contracts towards a normal Pop III star. To make this dependence explicit, let $M_{\rm AC,end}$ denote the
stellar mass at which AC support terminates.  For a given extended-AC
track we define the largest helium splitting that has become
kinematically accessible by that point as
\begin{equation}
    \delta_{\rm pre-dep}(M_{\rm AC,end})
    \equiv
    \max_{M_\star\leq M_{\rm AC,end}}
    \left[
        \eta_3 \mu_{\chi{\rm He}}
        \frac{G M_\star}{R_\star(M_\star)}
        \right].
    \label{eq:delta_predep}
\end{equation}
A splitting $\delta<\delta_{\rm pre-dep}$ therefore opens at some
point while the star is still on the AC-supported branch, whereas
$\delta>\delta_{\rm pre-dep}$ requires additional contraction after
that branch terminates. As shown in Figure~\ref{fig:ac_accessibility}, a larger mass splitting requires a larger mass to reach the capture onset. For a mass splitting $\delta$ of dozens to hundreds of keV, a $>10^4\Msun$ dark star can reach the helium threshold while still on the extended-AC branch. Increasing the mass of the dark matter particle looses the requirement on the stellar mass. Small minihalos are also easier to reach the threshold than large minihalos, since the former are more compact at a given mass.

If AC
support ends before reaching the kinematic threshold, the inelastic channel may still turn on in the subsequent stellar contraction. This is dynamically different from the evolution studied here, because both the residual AC density and the time available to
build a captured reservoir may then be substantially reduced. Unless the inelastic capture channel provides substantial additional heating to stop the contraction, the star will continue to shrink until it becomes a normal Pop III star. The capture phycics of this scenario is beyond the scope of this work.

Having located the kinematic opening along the stellar growth tracks, we now follow the phase space populated by first capture.

\section{Post-capture phase-space evolution}
\label{sec:postcapture}

Crossing the compactness threshold turn on the first-scatter capture, but the captured particles can remain isolated from the AC component in both configuration and phase space. For understanding the subsequent evolution, we trace the post-capture distribution in specific orbital energy $E$ and specific angular momentum $L$, following the treatment of inelastic solor capture \cite{Blennow2018} with the solar potential and composition replaced by those of an $n=3$ primordial dark star.

\subsection{First-capture distribution}
\label{sec:firstcapture}

For the following orbital calculations we take $m_\chi=1.1$ TeV, $\delta=350$ keV, and a Maxwellian halo with one-dimensional dispersion $\sigma_u=10~{\rm km\,s^{-1}}$; the corresponding 566-keV distribution is quoted in Sec.~\ref{sec:rates}. The helium targets are approximated as stationary, and the recoil is drawn in the point-nucleus limit; Sec.~\ref{sec:rates} states where the empirical helium weak form factor of Sec.~\ref{sec:heliumresponse} is used instead. The incoming angular momentum is drawn from the focused halo flux and the recoil energy from Eq.~(\ref{eq:recoil_bounds}), with a uniform outgoing azimuth.  Energy and momentum conservation then determine the post-scatter orbit. The numerical sampling procedure is detailed in Appendix~\ref{app:kernels}.

Each captured particle is thereafter labelled by its specific orbital energy and angular momentum $(E,L)$. Its turning points $r_p$ and $r_a$ follow from
\begin{equation}
    E=\phi(r)+\frac{L^2}{2r^2},
    \label{eq:turningpoints}
\end{equation}
its radial speed is
\begin{equation}
    v_r(r)=\left[2\left(E-\phi(r)\right)-\frac{L^2}{r^2}\right]^{1/2},
    \label{eq:vr}
\end{equation}
and its radial period is
\begin{equation}
    P(E,L)=2\int_{r_p}^{r_a}\frac{dr}{v_r(r)} .
    \label{eq:period}
\end{equation}
These three integrals carry all of the orbital information used below.

The sampling shows that the first-capture orbits are highly nonthermal. Table~\ref{tab:firstorbits} summarizes the orbital distribution at three values of the stellar compactness. The typical binding energy decreases from $\sim 3\delta$ at $\lambda=0.5$ and converges to $\sim 2\delta$ as the star approaches the threshold. This is a consequence of the shrinking kinematically allowed interval $E_{\pm}\rightarrow \frac{\mu_{\chi{\rm He}}\delta}{m_{\rm He}}\simeq \delta$ near the threshold, given Eq.~\ref{eq:Eafterup}. Compared with the typical depth of the central potential of TeV-mass dark matter
\begin{equation}
    \frac{GM_\star m_{\chi}}{R_\star}\sim \frac{m_{\chi}}{\mu_{\chi{\rm He}}}\frac{\delta}{\lambda}\simeq 10^2 \delta,
\end{equation}
the first-capture orbits are weakly bound at $\sim 1\%$ of the central potential depth, far from thermalization.

\begin{table}[!tbp]
    \caption{Properties of first-capture orbits for the 350-keV case, in the point-nucleus limit. Here $B\equiv-m_\chi E>0$, where $E$ is the specific orbital energy.}
    \label{tab:firstorbits}
    \begin{ruledtabular}
        \begin{tabular}{cccc}
            $R/R_{\rm kin,He}$ & ${\rm med}(B/\delta)$ &
            ${\rm med}(r_p/R)$ & ${\rm med}(r_a/R)$                   \\
            \hline
            0.99               & 2.00                  & 0.022 & 33.8 \\
            0.90               & 2.08                  & 0.071 & 36.2 \\
            0.50               & 2.95                  & 0.155 & 45.2
        \end{tabular}
    \end{ruledtabular}
\end{table}

The orbital geometry tells the same story. The median apocenter lies tens of stellar radii outside the photosphere even when the pericenter is deeply inside the star, indicating highly eccentric orbits. Even when the first-capture rate is large, a captured particle can spend most of its orbital period far outside the star. Effective thermalization therefore requires efficient subsequent scattering on the orbits that follow.

The spatial distribution of the DM population can also be derived from orbital integrals. Averaging
\begin{equation}
    p(r|E,L)=\frac{2}{4\pi r^2P(E,L)\,|v_r(r)|}
    \label{eq:pofr}
\end{equation}
over the orbits of an ensemble, with a normalization factor of  particle number, gives its orbit-averaged spatial density. Because overlap far outside the photosphere is irrelevant to the heating of the star, we characterize an ensemble by the stellar-interior coefficient
\begin{equation}
    A_\star\equiv\int_{r<R_\star}p^2(\bm r)\,d^3r,
    \qquad
    \frac{V_{{\rm eff},\star}}{R_\star^3}=\left[R_\star^3A_\star\right]^{-1},
    \label{eq:Astar}
\end{equation}
together with the interior fraction $f_{{\rm ann},\star}=A_\star/\int p^2\,d^3r$ and the single-particle residence fraction $f_{t,\star}$ inside the star. The residence quadrature and shell construction are given in Appendix~\ref{app:kernels}.

The first-capture orbital parameters, as shown in Table~\ref{tab:firstorbits}, have a weak dependence on the mass splitting. Raising $\delta$ from $350$ keV to the solar-bound $566$ keV changes the median orbital parameters by less than $5\%$ and leaves $f_{{\rm ann},\star}=0.4$ and $0.5$ at $\lambda=0.50$ and $0.90$ \footnote{Two weightings of the nuclear recoil appear here. Table~\ref{tab:firstorbits}, Table~\ref{tab:chain} and the transition times of this section use the point-nucleus contact interaction, for which $d\sigma/dE_R$ is flat; the terminal ensembles of Sec.~\ref{sec:terminal} carry the empirical helium form factor of Sec.~\ref{sec:heliumresponse}. Resampling first capture with the latter moves the median $r_a/R_\star$ from $45$ to $47$ at $\lambda=0.50$ and by less than $1\%$ at $\lambda=0.90$, and shifts $f_{{\rm ann},\star}$ by less than $1\%$ at both; at the stationary threshold the two weightings coincide identically, because the allowed recoil interval collapses to a point and a constant form factor drops out of the normalized recoil distribution. Appendix~\ref{app:kernels} carries the full finite-size dependence.}. At fixed $\lambda$ the reduced orbits depend on $\lambda$ alone, while $R_{\rm kin}\propto\delta^{-1}$ fixes the physical stellar scale and hence the target density. Only the timescales change with mass splitting.

The first-capture ensemble is thus fixed by the compactness alone, up to an overall timescale. What becomes of it depends on the state the particle is left in and on how long it waits for the next collision; we take these in turn.

\subsection{Excited state lifetime and post-capture scattering histories}
\label{sec:lifetime}

Right after the first capture, the DM particle is in the excited state $\chi_2$. The allowed next scattering depends on the comparison of the excited state lifetime with the next collision timescale, which is also modulated by the orbital period, given the highly eccentric orbits.

We first specify the post-capture state history for the prompt-decay Higgsino benchmark. At sub-GeV splittings the heavier neutral state decays predominantly through
\begin{equation}
    \chi_2\rightarrow\chi_1\gamma,
\end{equation}
via a $W$ loop \cite{GrahamRamaniWong2025}. Its decay length is well approximated by
\begin{equation}
    \ell_{\chi_2}
    \simeq7.46~{\rm km}
    \left(\frac{v}{400~{\rm km\,s^{-1}}}\right)
    \left(\frac{400~{\rm keV}}{\delta}\right)^3 ,
    \label{eq:higgsinoDecayLength}
\end{equation}
corresponding to
\begin{equation}
    \tau_{\chi_2}\simeq2.8\times10^{-2}~{\rm s}
\end{equation}
at $\delta=350$ keV. The quantity that matters is the ratio of this lifetime to the time to the next nuclear collision, which Sec.~\ref{sec:rates} finds to be of order $10^5$ yr, or $3\times10^{12}$ s: the Higgsino sits fourteen orders of magnitude inside the prompt-decay limit. As for the energy budget, the photon recoil is negligible for a TeV parent. We therefore take the ambient Higgsinos to be in $\chi_1$: without a process that continually repopulates the excited state, any primordial $\chi_2$ abundance decays away on negligible halo timescales. Thus every subsequent nuclear collision begins from the ground state.

A long-lived excited state instead permits a nuclear down-scatter before radiative return. Suppose a halo ground state particle is first captured through an endothermic transition and the captured excited state survives to undergo such a down-scatter. Its orbital energy immediately after the first collision is
\begin{equation}
    m_\chi E_1=
    \frac12m_\chi u^2
    -E_R^\uparrow-\delta.
    \label{eq:Eafterup}
\end{equation}
If the captured excited state later undergoes a down-scatter, then
\begin{align}
    m_\chi E_2
     & =m_\chi E_1-E_R^\downarrow+\delta \nonumber \\
     & =\frac12m_\chi u^2
    -E_R^\uparrow-E_R^\downarrow.
    \label{eq:updowncancel}
\end{align}
The splitting therefore cancels exactly after one complete up/down pair. In a primordial minihalo the incident kinetic energy is small compared with the characteristic recoil energies of the scattering events. A ground state halo particle is consequently difficult to re-eject in its first down-scatter.

For the contact interaction in the point-nucleus limit, constant $d\sigma/dE_R$ results in a uniform distribution of recoil energies over the kinematically allowed interval. The corresponding mean recoil in an exothermic down-scatter is
\begin{equation}
    \left\langle E_R^\downarrow\right\rangle
    =
    \frac{\mu_{\chi A}^2}{m_A}w^2
    +
    \frac{\mu_{\chi A}}{m_A}\delta.
    \label{eq:meanERdown}
\end{equation}
Substituting into the expression for the average energy change of the DM particle
\begin{equation}
    \left\langle\Delta E_\chi^\downarrow\right\rangle
    \equiv
    \left\langle m_\chi(E_2-E_1)\right\rangle
    =
    \delta-\left\langle E_R^\downarrow\right\rangle,
    \label{eq:meanEdm}
\end{equation}
we find
\begin{equation}
    \left\langle\Delta E_\chi^\downarrow\right\rangle
    =
    \left(1-\frac{\mu_{\chi A}}{m_A}\right)\delta
    -
    \frac{\mu_{\chi A}^2}{m_A}w^2.
    \label{eq:deltaEdown}
\end{equation}
For TeV-scale DM on helium, $\mu_{\chi A}/m_A\simeq 1$ drives energy loss over the relevant orbital velocities. Down-scattering therefore typically makes the orbit more tightly bound rather than promoting evaporation. The orbit integrations described in Appendix~\ref{app:kernels} bear this out: none of the $1{,}200$ simulated down-scattering orbits becomes unbound for the incident ground state halo particles considered here.

The excited state lifetime thus changes the sequence of subsequent collisions: prompt decay returns the Higgsino to $\chi_1$ before the next nuclear encounter, whereas a long-lived state permits alternating endothermic and exothermic transitions. These are the two limits of $\tau_{\chi_2}/t_{\rm scat}$, and for the Higgsino the intermediate regime is out of reach: Eq.~(\ref{eq:higgsinoDecayLength}) gives $\tau_{\chi_2}\propto\delta^{-3}$, so $\tau_{\chi_2}\sim t_{\rm scat}$ would require $\delta\lesssim10^2$ eV, more than three orders of magnitude below the splittings the LZ candidate motivates. We therefore treat the long-lived case not as the Higgsino at another splitting but as a stand-in for inelastic models whose excited state survives to scatter. Neither the radiative decay nor the long-lived down scattering provides an efficient evaporation channel. The two scenarios do, however, differ in how fast the next collision arrives, which we quantify next.

\subsection{Orbit-averaged transition rate and the first transition}
\label{sec:rates}

A captured particle scatters again when it next meets a nucleus inside the star. With the radial speed and orbital period of Sec.~\ref{sec:firstcapture}, the scattering optical depth per orbit in the stationary-nucleus and optically thin limits is
\begin{equation}
    \tau_A(E,L)=
    2\int_{r_p}^{\min(r_a,R_\star)}
    n_A(r)\,
    \sigma_A[w(r)]
    \frac{w(r)}
    {v_r(r)}\,dr .
    \label{eq:tauorbit}
\end{equation}
The mean waiting time is consequently
\begin{equation}
    t_{\rm scat}(E,L)=\frac{P(E,L)}{\tau_A(E,L)},
    \label{eq:tscat}
\end{equation}
and the orbit averaged transition rate is
\begin{equation}
    \Gamma_A(E,L)\equiv\frac{1}{t_{\rm scat}(E,L)}
    =\frac{\tau_A(E,L)}{P(E,L)} .
    \label{eq:Gamma}
\end{equation}

At fixed $\lambda\equiv{R_\star}/{R_{\rm kin,A}}$,
one has $R_\star\propto M_\star$. The density therefore scales as $\rho_\star\propto M_\star^{-2}$, the optical depth per orbit as $\tau_A\propto\sigma_A/M_\star$, and the orbital period as $P\propto M_\star$. Hence
\begin{equation}
    t_{\rm scat}\propto\frac{M_\star^2}{\sigma_A}.
    \label{eq:M2scaling}
\end{equation}
The steep $M_\star^2/\sigma_A$ scaling therefore makes post-capture relaxation increasingly sensitive to the scattering strength as the star grows. All times quoted below use $M_\star=3\times10^4M_\odot$ and $\sigma_{\rm He}^0=3.98\times10^{-37}\,{\rm cm^2}$ and may be rescaled with Eq.~(\ref{eq:M2scaling}).

For both prompt-decay and long-lived scenarios, the scattering obeys similar microphysics. Their major difference, the state of the captured particle, enters only through $\sigma_A$. In the point-nucleus limit the two contact cross sections are
\begin{equation}
    \sigma_\pm(w)=\sigma^0_{\rm He}\,\beta_\pm(w),
    \qquad
    \beta_\pm(w)=\left(1\mp\frac{2\delta}{\mu_{\chi{\rm He}}w^2}\right)^{1/2},
    \label{eq:sigmapm}
\end{equation}
with $\sigma_+$ for the endothermic transition $\chi_1\rightarrow\chi_2$ and $\sigma_-$ for the exothermic return $\chi_2\rightarrow\chi_1$. The prompt-decay scenario only allows the endothermic channel, whereas the long-lived excited state scenario requires both. The two cross sections are equal at high speeds, but the endothermic channel is suppressed near threshold: $\beta_+$ is real only for $w>v_{\rm th,He}$. In the following computations, we use the same coupling $\sigma^0_{\rm He}$ for both channels.

Putting the scattering microphysics in terms of the star potential, we find that the velocity-dependent factors in Eq.~(\ref{eq:sigmapm}) depends on the stellar compactness and the local escape speed through
\begin{equation}
    \beta_\pm=\left[1\mp\frac{\lambda}{g(r)}\right]^{\frac{1}{2}}, \quad g=\frac{v_{\rm esc}^2(r)}{v_{\rm esc}^2(0)},
\end{equation}
so, at fixed compactness neither cross section depends on the splitting. At a given relative speed, the exothermic cross section is the larger one, by
\begin{equation}
    \frac{\beta_-}{\beta_+}=\left(\frac{1+\lambda/g}{1-\lambda/g}\right)^{1/2},
    \label{eq:betaratio}
\end{equation}
which at the stellar center is $1.1$, $1.7$, $4.4$ and $14$ at $\lambda=0.10$,
$0.50$, $0.90$ and $0.99$. The threshold acts a second time through the geometry:
the endothermic channel is confined to the region $g(r)>\lambda$, a central sphere
that shrinks to nothing as $\lambda\rightarrow1$, whereas the exothermic one is
allowed anywhere inside the star. Section~\ref{sec:stall} applies Eq.~(\ref{eq:Gamma}) collision by collision to simulate the post-capture evolution with rescattering.

\begin{figure}[!tbp]
    \centering
    \includegraphics[width=\columnwidth]{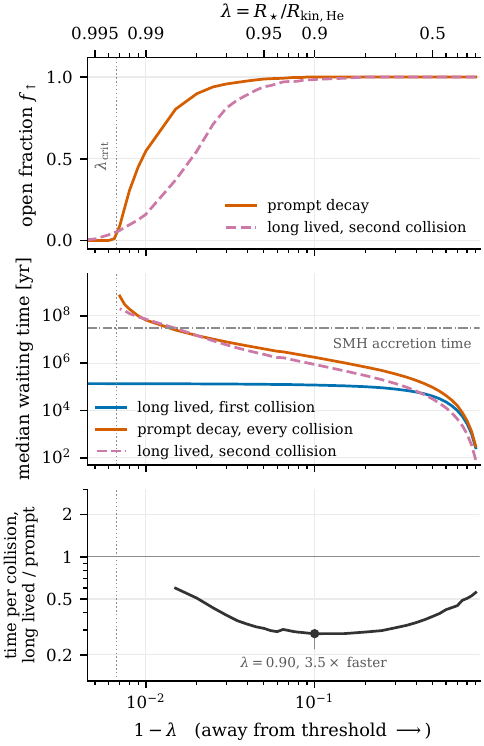}
    \caption{Post-capture relaxation clock for the two excited state lifetimes, at
        $\delta=350$ keV in the point-nucleus limit, from $400$ first-capture orbits per
        compactness and $2000$ for $\lambda\ge0.94$. Upper panel: the fraction of the sample
        for which the next collision is kinematically open. Under prompt decay this is a
        ground state up-scatter from the first-capture orbit and applies at every collision;
        for a long-lived state it is the up-scatter at its second collision, after one
        exothermic return has tightened the orbit. Middle panel: median waiting times over
        the open subset for $M_\star=3\times10^4M_\odot$ and
        $\sigma_{\rm He}^0=3.98\times10^{-37}\,{\rm cm^2}$, scaling as
        $M_\star^2/\sigma_{\rm He}$; a long-lived state alternates between its two curves
        while prompt decay repeats the middle one. The dash-dotted line marks the SMH
        accretion time at that stellar mass. Lower panel: the mean time a branch spends per
        collision, $\tfrac12(t_\downarrow+t_\uparrow')$ for a long-lived state against
        $t_\uparrow$ for prompt decay, plotted only where both subsets exceed $30\%$ of the
        sample. The dotted vertical line is $\lambda_{\rm crit}$ of
        Eq.~(\ref{eq:lamcrit}), beyond which neither branch can scatter at all.}
    \label{fig:branchscan}
\end{figure}

Instead of the physics of a single collision, the excited state lifetime controls the clock of the whole relaxation. The two scenarios of Sec.~\ref{sec:lifetime} set two different clocks. Under prompt decay, the particle is returned to $\chi_1$ on the first-capture orbit itself, so every one of its collisions is threshold limited: it must wait for an allowed up-scatter, and it must do so again after each one. A long-lived state instead reaches its next collision as $\chi_2$, for which $\beta_-$ in Eq.~(\ref{eq:sigmapm}) imposes no threshold at all; only at its second collision does it meet the threshold, and by then Eq.~(\ref{eq:deltaEdown}) has left the orbit more tightly bound. The relaxation then advances one collision at a time in the first case and in unthrottled--throttled pairs in the second. Figure~\ref{fig:branchscan} compares the two clocks.

The middle panel shows where the difference comes from. A long-lived state's first
collision is essentially free of the threshold: its waiting time changes by only a
factor of three between $\lambda=0.99$ and $0.50$, from $1.3\times10^5$ to
$4.0\times10^4$ yr, simply tracking the rising stellar density. Every prompt-decay
collision instead falls by five orders of magnitude across the same interval and is
$4.9\times10^2$ times slower than the unthrottled step at $\lambda=0.99$, a factor
that decreases monotonically to $1.2$ at $\lambda=0.10$. The two limits are the two
terms identified above: at $\lambda=0.10$ the measured contrast is the phase-space
ratio of Eq.~(\ref{eq:betaratio}) and nothing else, while at $\lambda=0.99$ that
ratio supplies only $14$ of the $4.9\times10^2$ and the rest is the collapse of the
allowed region. Deep inside the star the
excited state lifetime therefore stops mattering altogether; the distinction between
the two models is a near-threshold phenomenon.

Between those limits it matters a great deal. The lower panel compares the two
clocks on a common footing, as the mean time a branch spends per collision: a
long-lived state needs $\tfrac12(t_\downarrow+t_\uparrow')$ against $t_\uparrow$ for
prompt decay. The ratio has a minimum of $0.28$ near $\lambda=0.9$, where a
long-lived excited state relaxes $3.5$ times faster, and returns toward unity at both
ends of the range. That speed is not free. The same exothermic return that shortens
the clock also tightens the orbit, and thereby removes part of the population from
the chain altogether; Sec.~\ref{sec:stallcriterion} makes that trade quantitative and
Sec.~\ref{sec:terminal} gives its net effect on the relaxation as a whole.

Two features of Figure~\ref{fig:branchscan} carry into those chains. First, the
endothermic waits are long in absolute terms near threshold: the $6.5\times10^7$ yr
at $\lambda=0.99$ exceeds the $3\times10^7$ yr SMH accretion time at the same stellar
mass, so a reservoir captured just past the opening cannot relax appreciably before
the star grows through it. Second, raising the splitting to $\delta=566$ keV leaves
every open fraction unchanged and shortens every wait by a factor $\simeq5$. This is
exact in the point-nucleus limit: at fixed $\lambda$ and $M_\star$,
$R_\star\propto\delta^{-1}$ gives $n_{\rm He}\propto\delta^{3}$ and orbital speeds
$\propto\delta^{1/2}$, while $\sigma_\pm$ is independent of $\delta$ by
Eq.~(\ref{eq:sigmapm}), so $t\propto\delta^{-7/2}$ and $(566/350)^{7/2}=5.4$. The
same factor reappears in Sec.~\ref{sec:terminal}.

\subsection{Kinematic interruption of orbital relaxation}
\label{sec:stallcriterion}

Avoiding re-ejection does not imply that repeated state-changing scattering drives the captured dark matter to a Maxwell--Boltzmann distribution. Consider a bound ground state orbit with
\begin{equation}
    e\equiv\frac{E}{GM_\star/R_\star}.
\end{equation}
The largest orbital speed is obtained at the center. Using $\phi_c=-\eta_3GM_\star/R_\star$ and $v_{\rm th}^2=2\eta_3\lambda GM_\star/R_\star$, an up-scatter is possible somewhere in the star only if
\begin{equation}
    e>e_{\rm stall}\equiv-\eta_3(1-\lambda).
    \label{eq:estall}
\end{equation}
Once a ground state particle reaches $e\le e_{\rm stall}$, the leading off-diagonal nuclear interaction has zero up-scatter rate in the stationary-target, fixed-potential calculation, leaving a nonthermal stalled population. This ground state criterion applies after every radiative return in the prompt-decay Higgsino scenario and whenever a long-lived excited state is returned to $\chi_1$ by a nuclear down-scatter. Note that Eq.~(\ref{eq:estall}) involves the compactness alone: it is independent of the mass splitting, of the scattering strength, and of the stellar mass.

Equation~(\ref{eq:estall}) also bounds where post-capture relaxation can occur at all. Writing it in terms of the binding energy $B=-m_\chi E$ of Table~\ref{tab:firstorbits}, a radial orbit satisfies $e>e_{\rm stall}$ only if
\begin{equation}
    \frac{B}{\delta}<\frac{m_\chi}{\mu_{\chi{\rm He}}}\,\frac{1-\lambda}{\lambda},
    \label{eq:Bstall}
\end{equation}
whereas first capture delivers $B\rightarrow2\delta$ at the opening. The two coincide at
\begin{equation}
    \lambda_{\rm crit}
    =\frac{m_\chi/\mu_{\chi{\rm He}}}{2+m_\chi/\mu_{\chi{\rm He}}}
    =0.9933 ,
    \label{eq:lamcrit}
\end{equation}
above which a newly captured particle already lies at or below the stall boundary and the off-diagonal channel is shut on arrival. The sampled ground state open fraction in Figure~\ref{fig:branchscan} bears this out: it falls to $0.011$ at $\lambda=0.9935$ and to zero by $\lambda=0.994$, the finite angular momentum of real orbits closing them slightly before the radial bound.

Below $\lambda_{\rm crit}$ the criterion still selects, and it selects differently in the two branches, because a long-lived state arrives at Eq.~(\ref{eq:Bstall}) with an orbit that an exothermic return has already tightened. The upper panel of Figure~\ref{fig:branchscan} compares the two. On the first-capture orbit, which is what prompt decay faces at every collision, the channel is open for the entire sample down to $\lambda\simeq0.92$ and for $90\%$ of it at $\lambda=0.98$. After one exothermic return it is already incomplete at $\lambda\simeq0.85$ and has fallen to $0.55$ at $\lambda=0.98$. A long-lived excited state therefore buys its faster clock by losing particles from the chain, and it loses them fastest just past threshold, where Eq.~(\ref{eq:Bstall}) leaves the least room between first capture and the stall. The exothermic step is stochastic, however, and Eq.~(\ref{eq:deltaEdown}) gives only its mean: a small tail emerges less bound rather than more, which is why a few orbits remain open at $\lambda=0.994$, where every first-capture orbit has already closed.

The remaining elastic channels are too slow to bypass this off-diagonal stall on stellar-evolution timescales. Varying the spin-independent loop contribution over the range quoted by Pospelov and Ramani while retaining their spin-dependent hydrogen contribution gives median elastic-relaxation times of $2.6$--$2.9\times10^{13}$ yr at $\lambda=0.50$ and $9.3$--$10.3\times10^{13}$ yr at $\lambda=0.90$ on the $\delta=350$ keV first-capture orbits, and $0.51$--$0.57\times10^{13}$ yr and $1.7$--$1.9\times10^{13}$ yr on the corresponding $\delta=566$ keV orbits \cite{PospelovRamani2026}. These times exceed the $10^2$--$10^3$ yr stellar-adjustment interval by more than nine orders of magnitude, so the evolution below is governed by the leading off-diagonal channel.

First capture therefore produces an extended nonthermal reservoir whose typical apocenter lies tens of stellar radii away. Further off-diagonal relaxation can compact this reservoir as the stellar potential evolves, but state-changing kinematics can interrupt that contraction before a thermal core is reached. What distinguishes the inelastic case is not merely that first capture is nonthermal, but that this interrupted relaxation can preserve the extended phase long enough for its spatial distribution to matter. How far the relaxation actually proceeds before the channel closes cannot be read off from a single transition, however, since each collision moves the orbit and therefore moves the stall boundary with it. Section~\ref{sec:stall} follows the complete chains.

\section{Rescattering chains and the terminal kinematic stall}
\label{sec:stall}

Section~\ref{sec:postcapture} established two facts about a single postcapture
transition: the waiting time follows from the orbit  averaged optical depth, and a
ground state orbit below $e_{\rm stall}$ has no allowed up scatter anywhere in the
star. Neither statement determines where a captured particle ends up, because every
allowed collision changes $(E,L)$ and therefore changes both the next waiting time
and the stall condition itself. We therefore follow complete chains of successive
collisions in a fixed potential, from first capture to the kinematic endpoint, and
then ask what stellar contraction does to that endpoint.

Two scattering histories bracket the possibilities. In the prompt-decay Higgsino
benchmark, each up scatter is followed almost immediately by a radiative return to
$\chi_1$, so every collision is endothermic. A long-lived excited state instead
allows endothermic and exothermic collisions to alternate. We treat both. The
qualitative outcome rapid orbital compaction, terminating in a kinematic stall
that a small contraction reopens is common to the two limits, so it does not
depend on which one the underlying model realizes.

\subsection{Transition chains}
\label{sec:chains}

The chains evaluate the transition rate from Eq.~(\ref{eq:Gamma}) separately for
every orbit, so each particle carries its own transition rate rather than inheriting
the difference between two conditional median passage times. On a
collision, its radius is drawn from the  Eq.~(\ref{eq:tauorbit}) and its
recoil from the allowed inelastic interval. We draw the time to the next collision from an exponential distribution set by these rates, retaining particles with no kinematically allowed transition as a stalled population. The resulting passage distribution is therefore unconditional. Stalled particles remain part of the total population rather than being removed from the sample. Because Eq.~(\ref{eq:M2scaling})
is preserved chain by chain, all waiting times below can be rescaled by
$M_\star^2/\sigma_{\rm He}$. Compaction is measured by the stellar interior
coefficient $A_\star$ of Eq.~(\ref{eq:Astar}) and the associated
$V_{{\rm eff},\star}$, $f_{{\rm ann},\star}$ and $f_{t,\star}$.

\subsection{Compaction and attrition at fixed compactness}
\label{sec:compaction}

\begin{figure}[!tbp]
    \centering
    \includegraphics[width=\columnwidth]{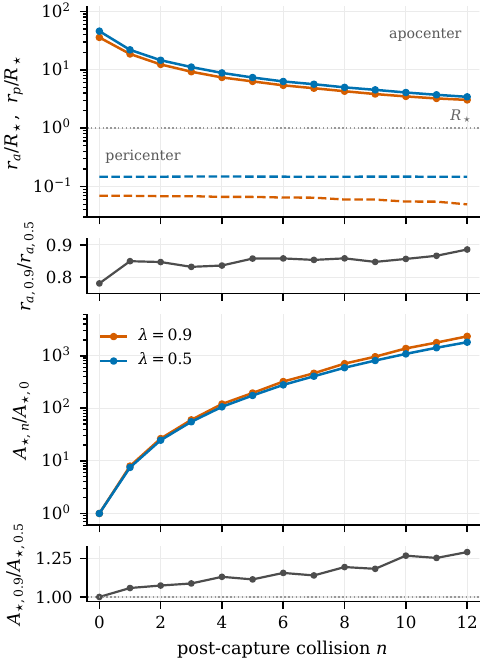}
    \caption{Orbital compaction of the captured population with successive collisions,
        for the alternating branch at $\delta=350$ keV in the point-nucleus limit, from $600$
        first-capture orbits at each compactness. First panel: median apocenter (solid, with
        markers) and median pericenter (dashed), in units of the stellar radius; the dotted
        line is the photosphere. Third panel: the stellar-interior coefficient $A_{\star,n}$
        of Eq.~(\ref{eq:Astar}), normalized to its first-capture value at the same
        compactness. The two narrow panels give the ratio of the $\lambda=0.90$ curve to the
        $\lambda=0.50$ one, which the logarithmic axes above compress; dotted lines mark
        unity. Tabulated values are given in Appendix~\ref{app:chaintable}.}
    \label{fig:chainevol}
\end{figure}

Figure~\ref{fig:chainevol} follows $600$ first captured orbits through twelve
successive collisions at fixed compactness, for the alternating (long-lived) branch
at $\delta=350$ keV. Appendix~\ref{app:chaintable} tabulates the same quantities,
together with the completion fractions and elapsed times, at the collision numbers
quoted below. Twelve is an integration cutoff here. Sec.~\ref{sec:terminal} follows the same chains to where no transition
remains allowed.

From the alternating branch scattering chains of the both compactnesses, we find three common features. First, relaxation removes the
extended outer part of the orbit rather than translating the whole orbit inward. The
first panel shows the two turning points moving in opposite ways: at $\lambda=0.90$
the median apocenter falls by an order of magnitude, from $35.8$ to $3.0R_\star$,
while the median pericenter changes only from $0.069$ to $0.050R_\star$. Even after
twelve collisions the apocenter is still several stellar radii out, so the population
has not become a core in the stellar interior.

Second, the compaction per collision barely depends on where the star sits relative
to threshold. This is what the two ratio panels resolve. The first collision erases
most of the compactness dependence of the apocenter, after which the two ensembles
keep a fixed offset: the median apocenter at $\lambda=0.90$ stays between $0.83$ and
$0.89$ of the value at $\lambda=0.50$ from $n=1$ onwards, having started at $0.78$.
The interior coefficient runs ahead at $\lambda=0.90$ by only $29\%$ after twelve
collisions, against the factor $2\times10^3$ by which each has grown. What the
compactness controls is therefore how long the collisions take, not what each one
does to the orbit. The interior fraction $f_{{\rm ann},\star}$ meanwhile rises only
from $0.49$ to $0.71$ at $\lambda=0.90$, so a substantial part of the overlap is
still accumulated outside the photosphere.

Third, the compaction is not free: at $\lambda=0.90$ only half of the initial orbits
complete twelve collisions, the remainder having stalled, whereas at $\lambda=0.50$
the chain is essentially loss-free over the same number of collisions
($f_{12}=0.998$).

The prompt-decay branch behaves the same way at the same splitting. Its
unconditional fractions at $\lambda=0.90$ are $1.00$, $0.99$, $0.93$, $0.77$ and
$0.46$ after $1$, $2$, $4$, $8$ and $12$ collisions, and its median apocenter falls
from $36.6$ to $2.7R_\star$ over the same interval. The attrition and the compaction
are therefore properties of the kinematics rather than of the excited state lifetime.

\subsection{Relaxation versus contraction}
\label{sec:actiontransport}

The rescattering chains hold the star potential fixed while the orbits evolve, but a real star contracts
on a comparable timescale as shown in Figure~\ref{fig:branchscan}. So, in an evolving calculation the growth of $A_\star$
would have two possible sources: the collisions themselves, and a deepening
potential dragging every bound orbit inward. To compare the contribution from the two, we transport each collision generation
from $\lambda=0.90$ to $\lambda=0.50$ at conserved $L$ and $J_r$.
Across the first twelve post-capture collisions the physical interior coefficient
satisfies
\begin{equation}
    0.85<\frac{A_{\star,n}(0.50)}{A_{\star,n}(0.90)}<1.03 ,
    \label{eq:D2bstagekernel}
\end{equation}
as shown in the upper panel of Figure~\ref{fig:D2bstall}. Contraction by a factor
$1.8$ in radius therefore leaves the population essentially where it was, against
the three orders of magnitude that the collisions supply. The reason is visible in
Figure~\ref{fig:chainevol}: these orbits spend most of their period outside the
photosphere. In these region, the Keplerian potential is conserved during the contraction at fixed
stellar mass. In units of the stellar radius the overlap $R_\star^3A_\star$ falls by
a factor of $7$ under the same transport, almost entirely because $R_\star^3$ has
shrunk by $5.8$. Instead of compacting the captured population, the star in fact withdraws from it.

\subsection{The terminal stall}
\label{sec:terminal}

How close that cutoff sits to the end depends strongly on the compactness. At
$\lambda=0.90$ the median number of completed post-capture collisions is $11$, with
a $16$th--$84$th percentile range of $7$--$15$ and a maximum of $23$: the twelfth collision is already close to the terminal distribution. At $\lambda=0.50$ the median is instead $97$, with a range of $65$--$113$ and a maximum of $135$, so Figure~\ref{fig:chainevol} there shows only the first eighth of the chain.

The middle panel of Figure~\ref{fig:D2bstall}
gives the full attrition curves, continued past the twelve collisions of
Figure~\ref{fig:chainevol} to the terminal stall. The extra collisions happen faster than the previous scattering due to their compacted later
orbits. For $M_\star=3\times10^4M_\odot$ and
$\sigma_{\rm He}^0=3.98\times10^{-37}\,{\rm cm^2}$ the median stochastic time to the
stall is $2.4$ Myr at $\lambda=0.90$ and $0.15$ Myr at $\lambda=0.50$; the latter
exceeds the median time to the twelfth collision by less than a factor of two despite
requiring roughly eighty additional collisions. All $600$ chains at each compactness
terminate in a kinematic stall, with no evaporation and no numerically invalid
transition.

The prompt-decay branch stalls after the same number of collisions but takes longer
to get there, since every transition must now be endothermic: at $\lambda=0.90$ the
median is again $11$ collisions ($6$--$14$, maximum $16$) but the median stall time is
$5.9$ Myr, and at $\lambda=0.50$ the median is $87$ collisions ($53$--$105$, maximum
$112$) and $0.32$ Myr. Table~\ref{tab:terminal} follows this branch over a wider range
of compactness. The stall is reached after roughly ten collisions near threshold and
after more than a hundred at $\lambda\le0.4$, but the time to reach it falls by more
than two orders of magnitude over the same interval, because both the collision rate
and the degree of binding increase as the star contracts. The terminal distribution
also changes character: at $\lambda=0.90$ the stalled orbits still have median
apocenters of about $3R_\star$ and spend only $10\%$ of their time inside the star,
whereas by $\lambda=0.50$ they are contained within the star, spend $93\%$ of their
time inside it, and accumulate essentially all of their overlap there. The stalled
population is thus extended and only partially star-crossing when the channel first
opens, and compact only well above threshold.

Repeating the same calculation at the solar-bound splitting $\delta=566$ keV leaves
every entry of Table~\ref{tab:terminal} unchanged to within $1\%$ except the
timescales, which shorten by a factor $\simeq5$, to $1.14$, $0.230$, $0.062$ and
$0.029$ Myr at $\lambda=0.90$, $0.70$, $0.50$ and $0.40$. This is the $t\propto\delta^{-7/2}$ scaling
derived in Sec.~\ref{sec:rates}: the stall condition and the reduced orbits depend
only on $\lambda$, and the splitting enters solely through the physical stellar
radius.

\begin{table}[!tbp]
    \caption{Terminal fixed potential ensembles for the prompt-decay Higgsino at
        $\delta=350$ keV, computed with the empirical helium form factor, from $150$
        first-capture orbits per compactness. Here
        $n_{\rm stall}$ is the number of completed post-capture collisions,
        $t_{\rm stall}$ is the median stochastic time for $M_\star=3\times10^4M_\odot$ and
        $\sigma_{\rm He}^0=3.98\times10^{-37}\,{\rm cm^2}$, and $f_{t,\star}$ is the
        single particle residence fraction inside the star; a tilde denotes the ensemble
        median. At $\lambda=0.30$ and $0.25$ only $36\%$ and $13\%$ of the chains stall
        before the $160$ collisions integration cap, so those two rows are lower bounds.}
    \label{tab:terminal}
    \begin{ruledtabular}
        \setlength{\tabcolsep}{3pt}
        \begin{tabular}{ccccccc}
            $\lambda$                       & $\tilde n_{\rm stall}$ & $t_{\rm stall}$ [Myr] &
            $\tilde r_a/R_\star$            & $f_{t,\star}$          &
            $V_{{\rm eff},\star}/R_\star^3$ & $f_{{\rm ann},\star}$                                                            \\
            \hline
            0.90                            & 11                     & 5.91                  & 2.98 & 0.096 & 93   & 0.714     \\
            0.70                            & 42                     & 1.12                  & 0.96 & 0.736 & 3.9  & 0.987     \\
            0.50                            & 87                     & 0.319                 & 0.54 & 0.929 & 0.93 & 0.9997    \\
            0.40                            & 125                    & 0.136                 & 0.41 & 0.970 & 0.45 & 0.9999    \\
            0.30                            & $>160$                 & $>0.048$              & 0.35 & 0.990 & 0.27 & $>0.9999$ \\
            0.25                            & $>160$                 & $>0.025$              & 0.36 & 1.00  & 0.27 & 1.0000
        \end{tabular}
    \end{ruledtabular}
\end{table}

Relative to first capture, the terminal ensembles have
$A_{\star,{\rm stall}}/A_{\star,0}=2.0\times10^3$, $9.3\times10^4$, $6.7\times10^5$
and $1.6\times10^6$ at $\lambda=0.90$, $0.70$, $0.50$ and $0.40$. They are far more compact than the first-capture distribution. The final level of compaction depends on the star compactness relative to threshold. But the distribution of orbital parameters at the stall is still far from a inside-the-star thermal core which appears at the elastic capture.

\subsection{Reopening of the inelastic channel}
\label{sec:reopening}

\begin{figure}[!tbp]
    \centering
    \includegraphics[width=\columnwidth]{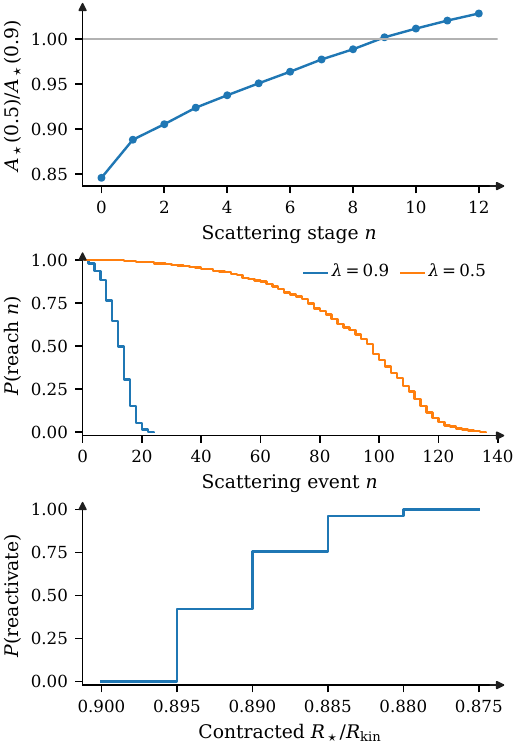}
    \caption{Action transport and kinematic stall of the metastable chain. The
        upper panel gives the physical stellar-interior annihilation coefficient
        after transporting the populations produced after each successive scatter from $\lambda=0.90$ to $\lambda=0.50$. The middle
        panel gives the unconditional fraction of particles that complete each successive scatter at fixed
        compactness. The lower panel follows the $\lambda=0.90$ stalled population
        under contraction; all of these ground state orbits regain an allowed
        upscatter by $\lambda=0.88$. The staircase appearance reflects evaluation at intervals $\Delta\lambda=0.005$.}
    \label{fig:D2bstall}
\end{figure}

A kinematic stall under fixed potential can be broken under stellar contraction. To examine this, we transport the
stalled ground state orbits to smaller radii at conserved actions and recheck the
pericenter speed. The results show that the endothermic channel reopens after a very small
contraction. For the alternating ensemble stalled at $\lambda=0.90$, $42.2\%$ of the orbits have
reopened by $\lambda=0.895$, $75.7\%$ by $0.890$, $96.2\%$ by $0.885$, and all $600$
by $0.880$; the median reopening point, evaluated on a grid of spacing
$\Delta\lambda=0.005$, is $0.890$. The lower panel of Figure~\ref{fig:D2bstall} shows
this reopening. The prompt-decay ensembles reopen at the same place and also away
from threshold: every orbit stalled at $\lambda=0.90$, $0.70$ and $0.50$ regains an
allowed up-scatter by $\lambda=0.884$, $0.689$ and $0.494$, apart from a single orbit
at $\lambda=0.50$ that requires $0.488$, corresponding to radius contractions of
$1.8\%$, $1.6\%$ and $1.3\%$ ($2.5\%$ for that one orbit). Because the stall
condition of Eq.~(\ref{eq:estall}) involves only $\lambda$, the reopening grid is
identical at $\delta=350$ and $566$ keV.

A percent level contraction therefore returns a stalled population to the collisional relaxation. Whether this matters depends on the ordering of two timescales. The
median formation time of the stall itself is $2.4$ Myr at $\lambda=0.90$, far longer
than the time a fixed mass star of this compactness would take to contract through
the same interval without additional support. Following repeated stall and reopening
over stellar evolution times therefore requires the potential and the phase space to
be evolved together, which is beyond the fixed potential treatment used here.

The picture that emerges from this section is a reservoir that neither thermalizes
nor stays at the first capture orbits. It compacts by three orders of magnitude in
interior overlap within roughly a dozen collisions. It stalls at eccentric orbits that are still only partly star crossing. The contraction from star evolution can restart the relaxation. The population that contributes to
annihilation heating is therefore set by the stellar evolution rather than the scattering strength of captured particles.

\section{Discussion}
\label{sec:discussion}

\subsection{From nonthermal capture to a persistent reservoir}

The compactness threshold determines when endothermic capture becomes possible; the orbital distribution produced by that capture determines what the new channel can subsequently do. The kinematics that opens the capture channel then interrupts the relaxation when the specific orbital energy drops below another threshold $e_{\rm stall}$ set only by $\lambda$. So, if the stellar is in stationary, the captured population remains in an extended spatial configuration. However, as the accretion and contraction drives the star to higher $\lambda$, the stall condition is also crossed, so the captured population advances in bursts separated by stalls, rather than relaxing monotonically toward a thermal core.

Although not considered quantitatively in this work, the annihilation of the captured population can change the evolution of the post-capture reservoir dramatically. The annihilation rate depends on both the compactness of the reservoir (in the digonal channel) and the orbital overlap with the star (in the off-diagonal channel). Both channels are controlled by the kinematic evolution of the orbital distribution. On the other hand, the annihilation heating of the captured population can also change the stellar structure, which in turn breaks the assumed AC-supported evolution. This may avoid the burst-stall-burst scenario described above, but the details depend on the specific dark matter model and the stellar evolution.

\subsection{Relation to the conventional elastic capture picture}

Elastic capture also passes through a nonthermal, star-crossing stage before repeated scattering produces a compact distribution, and conventional capture-supported dark-star evolution treats that stage as a short thermalization transient. Our results do not overturn that treatment everywhere. Deeply inside the threshold, the inelastic chains behave much as the conventional picture assumes: below $\lambda\simeq0.8$ every chain in Figure~\ref{fig:branchscan} is loss-free, and the threshold suppression of the endothermic channel, a factor $4.9\times10^2$ at $\lambda=0.99$, has fallen to $7$ at $\lambda=0.80$ and to $1.2$ at $\lambda=0.10$. Because the radius fits behind Eq.~(\ref{eq:MkinScaling}) also give $\lambda\propto M_\star^{-0.87}$, this confines the departure to $M_{\rm kin}<M_\star\lesssim1.3\,M_{\rm kin}$: on the SMH track at the solar-bound splitting, from $5.4\times10^4$ to about $7\times10^4M_\odot$, an interval a growing star crosses within a third of its mass.

The nonthermality of inelastic capture is more than a timescale problem. The end of relaxation also differs from the convenional picture. Elastic relaxation is monotone and terminates when the captured particles reach the local temperature. The inelastic chains terminate at $e_{\rm stall}$, a boundary independent of the scale of the system. The boudary set by Eq.~\ref{eq:estall} outlines a region of orbital parameter space that is not allowed to escape. Particles that are trapped in this region are typically still spatially extended. The terminal here is kinematic rather than thermal, in the sense that it can be reactivated. A contraction of $1.3$--$1.8\%$ is enough to reopen the relaxation.

The model complexity of the inelastic captured population is much higher than the thermalized elastic population for evaluating the annihilation heating and the consequence on stellar evolution.
In the conventional picture the captured component has a fixed shape, so a single number specifies it and the annihilation rate follows from a thermal core radius.
The persistence of a nonthermal reservoir in eccentric orbits, on the other hand, allow rich interactions and mechanisms involving the captured-captured annihilation in the reservoir, the captured-AC annihilation when the reservoir enters the star, and the orbital evolution accompanied by these interactions.
So, it is then necessary to simulate the nonthermal reservoir and its orbit evolution together with successive scattering.

\subsection{Dependence on the dark matter model}

The Higgsino benchmark fixes the normalization of the nuclear
interaction. Other models dominated by endothermic scattering can
exhibit similar orbital evolution, although their quantitative
predictions need not agree. As long as the scattering is endothermic, the kinematic threshold and the stall condition will be present.

The lifetime of $\chi_2$ determines the sequence of nuclear
collisions. With prompt radiative decay, each captured particle
returns to $\chi_1$ before its next collision. When $\chi_2$
survives until that collision, endothermic and exothermic
transitions can alternate. For the scattering inputs used here,
both limits produce orbital compaction and kinematic stalls that
can be reopened by stellar contraction, although the times required
to reach the stall differ. Neither scenario provides efficient
ejection of particles captured from the incident ground state
population. No evaporation occurs in the simulated
downscattering histories. A persistent $\chi_2$ population could
also permit additional annihilation channels, including
coannihilation, whose contributions to the heating would depend
on the abundances of both states. In conclusion, the kinematic picture presented here are generic regardless of the relative lifetime of the excited state, but including the annihilation heating may result in different quantitative predictions.

Embedding this picture in a self-consistent dark-star sequence requires one further condition: the ground state dark matter must possess an efficient diagonal annihilation channel. This requirement is already implicit in the initial stellar model, since the AC-supported sequence assumes that annihilation of the pre-existing contracted dark matter supplies the stellar luminosity. In a prompt-decay model with negligible $\chi_1\chi_1$ annihilation, the kinematic threshold and an extended nonthermal captured reservoir could still occur, but the AC-supported dark star assumed here would not be realized in the first place.

\subsection{Scopes and limitations}

The assumptions we made for simplicity limit the interpretation of the results above.

The threshold and orbital calculations are performed in a fixed $n=3$ stellar potential, continued as $-GM_\star/r$ outside the star. Stellar contraction is treated by comparing calculations at different compactness rather than by evolving the potential and the phase space together; the reopening of a stalled population under contraction (Sec.~\ref{sec:reopening}) is obtained by transporting conserved actions between fixed potentials.

Target motion enters the capture rate and broadens the stationary threshold by an amount equivalent to a zero-temperature penetration $1-\lambda\simeq0.04$--$0.05$ at the reconstructed crossing temperatures. The first-capture orbital distribution is nevertheless drawn from stationary-target kinematics, so this thermal rounding is included in the capture onset but not propagated into the initial orbital ensemble. Once the channel is well open the thermal correction rapidly decreases, but a fully thermal treatment of the first-capture phase-space distribution would be required to resolve the narrow edge itself.

The reference $M_\star$--$R_\star$ relations are the $m_\chi=1000$ GeV, AC-supported SMH and LMH sequences of Ref.~\cite{RindlerDaller2015}, chosen because they closely match the $1.1$-TeV Higgsino benchmark used in the capture calculation. We have nevertheless not recomputed the stellar structure at exactly $1.1$ TeV. More importantly, these reference tracks omit scattering capture and the stellar response to the captured population. Due to the threshold feature, ignoring the capture is exact until the onset of inelastic capture. However, once capture heating becomes dynamically important, the prescribed AC-supported sequence is no longer self-consistent. A complete treatment requires real-time evolution of the annihilation heating and the stellar structure. Moreover, the model dependence of the complete treatment is stronger than our orbital calculation here.

The annilation and off-diagonal scattering of captured particles off the bound AC component are not included in this work. The major reason is that the relative strength between the annilation and the scattering has strong model dependence. But the self-scattering is unlikely to change the qualitative picture of the kinematic evolution. It is in any case subdominant to the helium channel as
a relaxation mechanism, since $n_{\rm AC}$ is smaller than $n_{\rm He}$ by some eight orders of magnitude for a typical AC-supported dark star.

We do not consider the halo-scale long-term kinematic evolution, Neither the Capture from bound star-crossing trajectories nor the loss cone effect \cite{Sivertsson2011} is important for the kinematic picture presented here. We separate the kinematic evolution relevant to the inelastic capture itself, living in a much shorter timescale compared to the long-term evolution of the halo-scale dark matter distribution.

\subsection{Conclusions}
\label{sec:conclusion}

Endothermic capture changes both when dark matter can be captured by a primordial dark star and how the capture population evolves thereafter. The main findings are:

\begin{itemize}
    \item Endothermic nuclear scattering converts capture into a condition on stellar compactness. For an $n=3$ dark star the channel first opens at the center when $R_\star<R_{\rm kin,A}$, with $R_{\rm kin}\propto \mu_\chi M_*/\delta$, and the rate turns on quadratically, $J_3\propto(1-\lambda)^2$. Every allowed first scatter is gravitationally binding over the relevant phase space.

    \item Thermal motion of the stellar targets rounds this opening without removing it: the central temperatures of the $m_\chi=1000$ GeV tracks give $T_c\simeq(0.86$--$1.08)\times10^7$ K at the $350$ keV crossings, corresponding to an equivalent zero temperature penetration $1-\lambda\simeq0.04$--$0.05$. The correction rapidly decreases once the channel is well open. Helium dominates the capture rate over the relevant compactness range, and the empirical helium weak form factor preserves the quadratic opening law while suppressing the normalization by $6\%$ at $\delta=350$ keV and $10\%$ at $566$ keV.

    \item For a $1.1$ TeV Higgsino the helium channel opens at $M_{\rm kin}\simeq5.8\times10^3\,M_\odot$ (SMH) and $9.2\times10^3\,M_\odot$ (LMH) for $\delta=350$ keV, moving to $1.1\times10^4$ and $1.6\times10^4\,M_\odot$ at the current solar bound $\delta>566$ keV. MeV-scale splittings, which lie beyond present solar reach, remain kinematically accessible at supermassive-dark-star masses.

    \item First capture is strongly nonthermal. Typical binding energies are of order $2\delta$ and median apocenters lie tens of stellar radii beyond the photosphere, so a captured particle spends only about $0.2\%$ of its orbital period inside the star.

    \item Neither of the two possible post-capture state scenarios, the prompt decay and the long-lived excitation state, provides an efficient evaporation channel. Prompt radiative return is effectively recoil free for TeV dark matter. An exothermic collision between a heavy excitation state $\chi_2$ and He is negative, so down-scattering also tightens the orbit.

    \item Repeated state changing scattering does not drive the captured population to a Maxwell--Boltzmann distribution. Below $e_{\rm stall}=-\eta_3(1-\lambda)$ the leading off-diagonal channel has no allowed up-scatter anywhere in the star, and the remaining elastic channels give relaxation times of $10^{13}$--$10^{14}$ yr and cannot bypass the stall. Every simulated chain therefore ends in a kinematic stall rather than in a thermal core, after a median of $11$ collisions at $\lambda=0.90$ and $\sim10^2$ at $\lambda\le0.50$.

    \item The chains compact the reservoir sharply but incompletely. Over twelve collisions the median apocenter falls from $36$ to $3R_\star$ and the stellar-interior annihilation coefficient rises by a factor $2\times10^3$, while the pericenter is nearly unchanged and only $71\%$ of the overlap is accumulated inside the star. Action transport shows this growth is a relaxation effect and not a geometric consequence of contraction, and the same behavior is obtained for prompt-decay and long-lived excited states scenarios.

    \item The stall is not a permanent endpoint. Transporting stalled orbits at conserved actions, a radius contraction of $1.3$--$1.8\%$ reopens the endothermic channel for the entire stalled population at every compactness examined. Because the stall forms on a Myr timescale, following repeated stall and reopening requires the stellar potential and the phase space to be evolved together.
\end{itemize}

The immediate implication is that captured dark matter in the inelastic case cannot be represented as an instantaneously thermalized fuel source. The reservoir remains extended and nonthermal over an interval set by the stellar compactness and the scattering strength, so its annihilation must be computed from the evolving orbital distribution. The heating that this reservoir supplies, and the stellar response to it along a supermassive-dark-star growth sequence, are developed in a companion paper.

\appendix

\section{Tabulated chain statistics}
\label{app:chaintable}

Table~\ref{tab:chain} collects the quantities plotted in Figure~\ref{fig:chainevol} and the middle panel of Figure~\ref{fig:D2bstall}, at the collision numbers quoted in Sec.~\ref{sec:compaction}.

\begin{table}[!tbp]
    \caption{Successive collision chains at fixed compactness for the alternating
        branch, $\delta=350$ keV, in the point-nucleus limit, from $600$ first-capture orbits per compactness. Here $n$
        counts post-capture collisions, $f_n$ is the unconditional fraction of the initial
        sample that completes the $n$th collision, and $t_n$ is the conditional median
        cumulative time for $M_\star=3\times10^4M_\odot$ and
        $\sigma_{\rm He}^0=3.98\times10^{-37}\,{\rm cm^2}$, scaling as
        $M_\star^2/\sigma_{\rm He}$.}
    \label{tab:chain}
    \begin{ruledtabular}
        \begin{tabular}{rcccccc}
            $n$                             & $f_n$                 & ${\rm med}(r_a/R_\star)$ & $A_{\star,n}/A_{\star,0}$ &
            $V_{{\rm eff},\star}/R_\star^3$ & $f_{{\rm ann},\star}$ & $t_n$ [Myr]                                                                           \\
            \hline
            \multicolumn{7}{c}{$\lambda=0.90$}                                                                                                              \\
            \hline
            0                               & 1.000                 & 35.8                     & 1                         & $1.9\times10^5$ & 0.49 & ---   \\
            1                               & 1.000                 & 18.6                     & 7.9                       & $2.4\times10^4$ & 0.56 & 0.12  \\
            2                               & 0.980                 & 12.3                     & 27                        & $7.3\times10^3$ & 0.57 & 0.89  \\
            4                               & 0.935                 & 7.4                      & $1.2\times10^2$           & $1.6\times10^3$ & 0.60 & 1.24  \\
            8                               & 0.765                 & 4.3                      & $7.1\times10^2$           & $2.8\times10^2$ & 0.65 & 1.59  \\
            12                              & 0.497                 & 3.0                      & $2.3\times10^3$           & 83              & 0.71 & 1.87  \\
            \hline
            \multicolumn{7}{c}{$\lambda=0.50$}                                                                                                              \\
            \hline
            0                               & 1.000                 & 45.9                     & 1                         & $5.0\times10^5$ & 0.40 & ---   \\
            1                               & 1.000                 & 21.9                     & 7.5                       & $6.7\times10^4$ & 0.42 & 0.035 \\
            2                               & 1.000                 & 14.6                     & 25                        & $2.0\times10^4$ & 0.43 & 0.068 \\
            4                               & 1.000                 & 8.8                      & $1.1\times10^2$           & $4.7\times10^3$ & 0.45 & 0.085 \\
            8                               & 1.000                 & 5.0                      & $5.9\times10^2$           & $8.5\times10^2$ & 0.48 & 0.099 \\
            12                              & 0.998                 & 3.4                      & $1.8\times10^3$           & $2.8\times10^2$ & 0.51 & 0.106
        \end{tabular}
    \end{ruledtabular}
\end{table}

\section{Numerical capture and orbital kernels}
\label{app:kernels}

This appendix gives the numerical procedures used for the threshold rounding, the first-capture orbital distribution, and the rescattering chains in Secs.~\ref{sec:captureonset}--\ref{sec:stall}. Monte Carlo quantities are quoted only to the precision supported by the sampling.

\subsection{Physical inputs and helium weak form factor}

We use $m_\chi=1.1$ TeV, $m_{\rm He}=3.727$ GeV, hydrogen and helium mass fractions $X=0.76$ and $Y=0.24$, and the $n=3$ density and potential of Sec.~\ref{sec:polytrope}, continued as $\Phi=-GM_\star/r$ outside the star. The orbital calculations in Secs.~\ref{sec:postcapture} and \ref{sec:stall} use $\delta=350$ keV for both the alternating and the prompt-decay branch; the solar-bound value $566$ keV is quoted where the splitting dependence is being tested. The helium cross section is $3.98\times10^{-37}\,{\rm cm^2}$ and $f_Q=2/3$. The halo has one-dimensional velocity dispersion $10~{\rm km\,s^{-1}}$. Prompt decay returns each captured particle to $\chi_1$ without appreciable recoil. Point-nucleus and long-lived comparisons are treated separately from this Higgsino prescription.

The $J^\pi=0^+$ helium ground state selects the coherent vector form factor. Writing $c_p=1-4\sin^2\theta_W$,
\begin{equation}
    F_W(q)=\frac{NF_n(q)-c_pZF_p(q)}{N-c_pZ},\qquad F_W(0)=1.
\end{equation}
Here the proton and neutron vector distributions include their nucleon electric structure. In the charge-symmetric, zero-strangeness one-body limit, the normalized weak and electromagnetic shapes of this isoscalar nucleus coincide \cite{Aniol2006}. We therefore use the empirical sum-of-Gaussians charge fit for ${}^4$He in Table V, p.~528, of de Vries \textit{et al.}\ \cite{DeVries1987} (entry Si82, measured over $q=0.14$--$7.70\,{\rm fm^{-1}}$). The fit has rms radius $1.676$ fm and individual Gaussian width $\gamma=\sqrt{2/3}$ fm. Its normalized Fourier transform is
\begin{equation}
    \begin{aligned}
        F_W(q)={} & e^{-q^2\gamma^2/4}
        \sum_i\frac{Q_i}{1+2R_i^2/\gamma^2}                                         \\
                  & \times\left[\cos(qR_i)+\frac{2R_i^2}{\gamma^2}j_0(qR_i)\right],
    \end{aligned}
    \label{eq:HeSOG}
\end{equation}
with $q$ in inverse femtometers and $\sum_iQ_i=1$. No additional nucleon-size factor multiplies this empirical form factor. Strange electric currents and isospin-breaking corrections are neglected in the helium form-factor calculation.

Table~\ref{tab:soginputs} gives the SOG coefficients. We use $\gamma=\sqrt{2/3}$ fm and normalize the tabulated $Q_i$ by their sum, which differs from unity by rounding. The momentum in Eq.~(\ref{eq:HeSOG}) is in fm$^{-1}$; $\hbar c=0.19732698~{\rm GeV\,fm}$. Replacing $F_W^2$ by unity gives the point-nucleus comparison.

\begin{table}[!tbp]
    \caption{Helium SOG inputs from the Si82 entry in Table V of Ref.~\cite{DeVries1987}. Paired columns continue the same list; $R_i$ is in fm.}
    \label{tab:soginputs}
    \begin{ruledtabular}
        \begin{tabular}{rrrr}
            $R_i$ & $Q_i$    & $R_i$ & $Q_i$    \\
            \hline
            0.2   & 0.034724 & 2.6   & 0.014201 \\
            0.6   & 0.430761 & 3.1   & 0        \\
            0.9   & 0.203166 & 3.5   & 0.006860 \\
            1.4   & 0.192986 & 4.2   & 0        \\
            1.9   & 0.083866 & 4.9   & 0.000438 \\
            2.3   & 0.033007 & 5.2   & 0
        \end{tabular}
    \end{ruledtabular}
\end{table}

\begin{figure}[!tbp]
    \centering
    \includegraphics[width=\columnwidth]{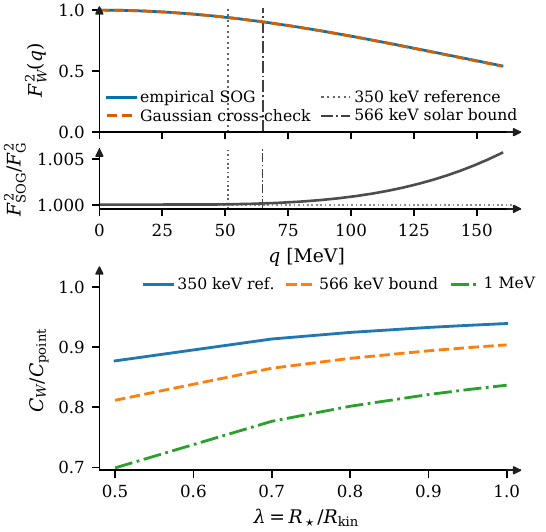}
    \caption{Helium weak form factor and finite-size correction. The upper panels compare the empirical
        SOG form factor with a radius-matched Gaussian, which is a sub-percent
        shape approximation over the relevant momentum range. The lower panel shows
        the finite-size suppression of the capture rate relative to a point nucleus
        for $\delta=350$ keV, 566 keV at the solar bound, and a 1-MeV splitting.
        Vertical marks denote the 350-keV and 566-keV threshold momenta.}
    \label{fig:A3response}
\end{figure}

Finite momentum transfer is retained inside each recoil integral. Figure~\ref{fig:A3response} summarizes the resulting finite-size correction. A radius-matched Gaussian reproduces the empirical SOG shape to sub-percent accuracy over the relevant momentum range; the larger physical correction is the finite-size suppression relative to a point nucleus. For $\delta=350$ keV and a halo dispersion of $10~{\rm km\,s^{-1}}$, the capture-rate ratio to the point-nucleus result is $0.877$, $0.925$, and $0.933$ at $\lambda=0.50$, $0.80$, and $0.90$, respectively.

At the stationary opening, the recoil interval collapses to $E_{R,\rm th}=\mu_{\chi\mathrm{He}}\delta/m_{\rm He}$, so $q_{\rm th}=\sqrt{2\mu_{\chi\mathrm{He}}\delta}=51.0$ MeV for $\delta=350$ keV. Since the form factor is smooth and nonzero there,
\begin{equation}
    J_{3,W}(\lambda)=2.10134 F_W^2(q_{\rm th})(1-\lambda)^2
    +\mathcal O[(1-\lambda)^3].
\end{equation}
The factor $F_W^2(q_{\rm th})=0.939$ at $\delta=350$ keV preserves the quadratic turn-on. At $\delta=566$ keV, $1$, $2$, and $5$ MeV, the same quantity is $0.904$, $0.837$, $0.701$, and $0.414$, respectively; the corresponding capture-rate ratios to the point-nucleus result at $\lambda=0.5$ are $0.812$, $0.699$, $0.511$, and $0.242$. The current solar Higgsino bound corresponds to $q_{\rm th}=64.9$ MeV on helium. For thermal targets, the same form factor enters the angular average in Eq.~(\ref{eq:thermalWeak}); at $\delta=350$ keV it reduces the rate at $\lambda=1$ by approximately $6.5\%$ for $T_c=5\times10^6$ K while retaining the few-percent compactness broadening. Raising the helium temperature from zero to $5\times10^6$ K changes the SOG capture ratio by less than $0.3\%$ at both compactnesses considered here, at $\delta=350$ keV ($0.24\%$) as well as at $566$ keV ($0.25\%$).

For the post-capture calculation, the same recoil weight determines the first-capture distribution and subsequent inelastic transitions. In particular,
\begin{equation}
    \sigma_\pm(w)=\frac{m_{\rm He}\sigma_{\rm He}^0}
    {2\mu_{\chi\mathrm{He}}^2w^2}
    \int_{E_-}^{E_+} F_W^2(\sqrt{2m_{\rm He}E_R})\,dE_R
\end{equation}
enters the orbit-averaged scattering rate, and the conditional recoil distribution is proportional to the same integrand. At $\lambda=0.50$, finite-$q$ weighting changes the median apocenter from about $45R_\star$ to $47R_\star$ and decreases the stellar-interior annihilation integral by $9.1\%$; at $\lambda=0.90$ both shifts are below $1\%$. In fixed-potential rapid-decay chains with $M_\star=3\times10^4M_\odot$ and $\sigma_{\rm He}^0=3.98\times10^{-37}\,{\rm cm^2}$, the median stall time at $\lambda=0.50$ changes from $0.27$ to $0.32$ Myr. Thus the finite nuclear-size correction modifies the relaxation rate without changing the extended, nonthermal character of the captured reservoir.

\subsection{Capture integration and sampling}

For stationary targets, the allowed recoil interval in Eq.~(\ref{eq:sigmacap}) is integrated including the capture lower bound. We use 128 Gauss--Legendre points over the allowed radial interval and 24 Gauss--Laguerre points in $u^2/(2\sigma_u^2)$, locating the outer allowed radius separately for each incoming speed. This resolves the shrinking central scattering volume near threshold instead of averaging over forbidden radii.

The thermal calculation uses the leading $u=0$ halo limit, $T(r)=T_c\theta(r)$, and Maxwellian target velocities. Three Gaussian target-velocity components and two angles for the outgoing relative velocity are sampled, with each collision weighted by $s\sqrt{1-v_{\rm th}^2/s^2}F_W^2$; binding is then evaluated in the stellar frame. The thermal integral is evaluated with a scrambled five-dimensional Sobol sequence using $2^{13}$ points and 128 radial points, with convergence checked using $2^{16}$ points, 256 radial points, and independent scrambles. The temperature comparison uses $T_c=0,2,5,7\times10^6$ K and $\lambda=0.5,0.9,1,1.03$. The resulting rate remains finite at the stationary threshold and receives only a modest correction once the channel is open.

For the orbit samples, we set $u=0$ and draw the collision radius $x=r/R_\star$ from
\begin{equation}
    dP\propto x^2\theta^3 g(x)\sqrt{1-\lambda/g(x)}
    \overline{F_W^2}(x)\,dx,
    \label{eq:app_injection}
\end{equation}
restricted to the allowed region. Here $\overline{F_W^2}$ is the average over the recoil interval. A cumulative distribution on 20001 uniformly spaced radii is used. The incoming $|v_r|/w$ and recoil azimuth are sampled uniformly, while the recoil energy is sampled with density proportional to $F_W^2(E_R)$. The orbital variables $(e,\ell)$ are obtained from the outgoing velocity, retaining only bound particles. The finite-$u$ correction belongs in the total capture rate; it is not a thermal correction to this orbit sample. About 2500 orbits at each of $\lambda=0.5,0.9$ reproduce median apocenters of several tens of stellar radii and an interior residence fraction near $2\times10^{-3}$.

\subsection{Residence probabilities and subsequent collisions}

Both turning points are found in the full interior-plus-exterior potential. The radial residence integral uses $x=(x_a+x_p)/2+(x_a-x_p)\sin\vartheta/2$ to remove endpoint singularities, with 96 Gauss--Legendre points in $\vartheta$. The normalized residence-time weights are deposited into radial shells. If $H_{ib}$ is the probability of orbit $i$ in shell $b$, its dimensionless density is $p_{ib}=H_{ib}/\Delta V_b$, with $\Delta V_b=4\pi x_b^2\Delta x_b$ evaluated at the geometric shell midpoint. All quadratic and linear overlaps are computed from the same densities, distinguishing the full orbital volume from shells entirely inside $x=1$. The radial mesh combines 160 logarithmic edges from $10^{-4}$ to 1 and 140 from 1.02 to 200; convergence is checked by varying both the orbit count and the shell mesh and by verifying probability coverage.

For each subsequent collision, the orbit-averaged rate is integrated over $x_p<x<\min(x_a,1)$ with the same finite-$q$ form-factor weighting. The scattering-radius quadrature uses 160 points after the transformation $x=x_p+(x_{\max}-x_p)s^2$. An exponential waiting time is sampled from that rate, followed by a collision radius from the normalized rate integrand and a recoil as above. Rapid decay permits only subsequent upscatters, whereas the long-lived comparison alternates the sign of $\delta$. A vanishing rate is treated as a physical stall rather than evaporation. Reach fractions are measured relative to the initial sample, and waiting-time medians are conditioned on reaching the specified collision. Samples of 400--600 chains reproduce the strong enhancement and the much longer waiting time near threshold. Chains are followed beyond twelve collisions when measuring terminal stall; the comparison calculation permits up to 220 collisions, well above the range relevant to the quoted stall statistics. At a different splitting, $R_{\rm kin}\propto\delta^{-1}$ is recomputed before reduced orbits are converted to physical velocities and rates.

The chain calculation was implemented twice, with different orbit counts, different random streams and an independent implementation of the transition sampling. At $\lambda=0.90$ the two give twelfth-collision completion fractions $0.497$ and $0.513$, interior enhancements $A_{\star,12}/A_{\star,0}=2.34\times10^3$ and $2.41\times10^3$, and median cumulative times agreeing to $8\%$. At $\lambda=0.50$ the enhancements are $1.81\times10^3$ and $2.09\times10^3$; the $13\%$ spread is the sampling error of the smaller, $160$-orbit ensemble, whose $A_\star$ is dominated by its few most compact orbits.

\bibliographystyle{apsrev4-2}
\bibliography{main}

\end{document}